\documentclass[pre,twocolumn,floatfix,showpacs]{revtex4-1}
\usepackage{graphicx}
\usepackage{dcolumn}
\usepackage{bm}
\usepackage{color}
\usepackage{natbib}
\usepackage{array}
\usepackage{subfigure}
\usepackage{amssymb}
\usepackage{amsbsy}
\usepackage{natbib}
\usepackage{epstopdf}
\usepackage{placeins}

\begin{document}

\title{Effect of Plumes on Measuring the Large Scale Circulation in Turbulent Rayleigh-B\'enard Convection}
\author{Richard J.A.M. Stevens$^1$}
\author{Herman J.H. Clercx$^{2,3}$}
\author{Detlef Lohse$^1$}
\affiliation{$^1$Physics of Fluids Group, Faculty of Science and Technology, Impact and MESA+ Institutes $\&$ Burgers Center for Fluid Dynamics, University of Twente, 7500AE Enschede, The Netherlands}
\affiliation{$^2$Department of Applied Mathematics, University of Twente, P.O Box 217, 7500 AE Enschede, The Netherlands}
\affiliation{$^3$Department of Physics and J.M. Burgers Centre for Fluid Dynamics, Eindhoven University of Technology, P.O. Box 513, 5600 MB Eindhoven, The Netherlands}
\date{\today}

\begin{abstract}
We studied the properties of the large-scale circulation (LSC) in turbulent Rayleigh-B\'enard (RB) convection by using results from direct numerical simulations in which we placed a large number of numerical probes close to the sidewall. The LSC orientation is determined by either a cosine or a polynomial fit to the azimuthal temperature or azimuthal vertical velocity profile measured with the probes. We study the LSC in $\Gamma=D/L=1/2$ and $\Gamma=1$ samples, where $D$ is the diameter and $L$ the height. For $Pr=6.4$ in an aspect ratio $\Gamma=1$ sample at $Ra=1\times10^8$ and $5\times10^8$ the obtained LSC orientation is the same, irrespective of whether the data of only $8$ or all $64$ probes per horizontal plane are considered. In a $\Gamma=1/2$ sample with $Pr=0.7$ at $Ra=1\times10^8$ the influence of plumes on the azimuthal temperature and azimuthal vertical velocity profiles is stronger. Due to passing plumes and/or the corner flow the apparent LSC orientation obtained using a cosine fit can result in a misinterpretation of the character of the large-scale flow. We introduce the relative LSC strength, which we define as the ratio between the energy in the first Fourier mode and the energy in all modes that can be determined from the azimuthal temperature and azimuthal vertical velocity profiles, to further quantify the large-scale flow. For $Ra=1\times10^8$ we find that this relative LSC strength is significantly lower in a $\Gamma=1/2$ sample than in a $\Gamma=1$ sample, reflecting that the LSC is much more pronounced in a $\Gamma=1$ sample than in a $\Gamma=1/2$ sample. The determination of the relative LSC strength can be applied directly to available experimental data to study high Rayleigh number thermal convection and rotating RB convection.
\end{abstract}


\maketitle

\section{Introduction} \label{sec_introduction}

Rayleigh-B\'enard (RB) convection is the motion of a fluid contained between two parallel plates which is heated from below and cooled from above \cite{ahl09,ahl09b,loh10}. A well established feature of the dynamics of the system is the large-scale circulation (LSC). It plays an important role in natural phenomena, including convection in the Arctic ocean, in Earth's outer core, in the interior of gaseous giant planets, and in the outer layer of the Sun.
The properties of the LSC have recently been intensively studied in experiments
\cite{nie01,xia03,fun04,sun05,sun05a,bro06,bro06b,xi07,bro07,bro08,bro08b,fun08,xi08,xi08c,kun08b,xi09,zho09,bro09,zho09b,ste09,sug10},
numerical simulations
\cite{ver03,str06,kun08b,kac09,mis11},
 and models
 \cite{sre02,ben05,fon05,res06,bro07,kun08b,bro08,bro08b,bro09,sug10}. For a complete overview of all literature in which certain aspect of the LSC are studied we refer to the recent review of Ahlers, Grossmann, and Lohse \cite{ahl09}.
 
In experiments the LSC is measured by using thermistors that are embedded in the sidewall, see e.g. \cite{bro05b}, or by using small thermistors that are placed in the flow at different azimuthal positions and different heights, see e.g. \cite{xi09}. Since the LSC transports warm (cold) fluid from the bottom (top) plate up (down) the side wall, the thermistors can detect the location of the up-flow (down-flow) by showing a relatively high (low) temperature. In addition, there have also been a number of direct measurements of the LSC by particle image velocimetry (PIV) and laser Doppler velocimetry (LDV) (see, for example Refs.\ \cite{xia03,sun05,xi06}) that complement the thermistor method.

In this paper we investigate the properties of the LSC with results from direct numerical simulations (DNS). Though DNSs are limited both in Ra and in duration, i.e. in number of LSC turnover times, to smaller values than the experimental analogues the advantage is that in contrast to the experiments the full spatial information is available. We will take advantage of this in order to verify the algorithms employed in experiments to identify the LSC orientation based on a limited number of probes. We will introduce two existing methods to determine the LSC orientation over time. The first method \cite{bro06} determines the LSC orientation from a cosine fit to the azimuthal temperature profile measured with the probes. This method has been extremely successful in revealing important features, i.e. azimuthal meandering, reversals, and cessations, of the LSC \cite{bro06,bro07,xi07,xi08c}. In the second method the LSC orientation is determined by using a second order polynomial fit around the maximum and the minimum. Using this method the sloshing mode of the LSC, which is caused by an off-center motion of the LSC, was discovered \cite{xi09,zho09,bro09}. We note that this can not be obtained by the cosine fit method as this method assumes that there is no off-center motion of the LSC.

So far these procedures have been applied to experimental data where the data of $8$ azimuthally equally spaced probes were available. The benefit of the simulations over the experiments is that it is easy to place a large number of probes in the flow. Here we placed up to $64$ azimuthally equally spaced numerical probes at different heights into the numerical RB sample. With an arrangement of up to $64$ probes we can determine how the extracted information on the LSC and plume dynamics depends on the number of probes that is used. We also visualized the flow by movies that show the flow dynamics in horizontal or vertical planes, see the accompanying material \cite{ste10movies}.

The paper is organized as follows. In section \ref{sec_numericalmethod} we start with a discussion of the numerical method that has been used. This is followed in section \ref{sec_LSCmethod} by a discussion of the traditional methods that are used in experiments to determine the LSC orientation. Based on the results obtained by these methods we discuss the characteristics of the LSC in a $\Gamma=1$ (section \ref{sec_Gamma1}) and $\Gamma=1/2$ (section \ref{sec_Gamma05}) sample. An important question for both high Ra number \cite{fun09,ste10,wei10} and rotating RB convection \cite{zho09b,ste09} is whether or not there is a single LSC present. Therefore we will discuss a new method in section \ref{sec_relativeLSCstrength},
which is based on the energy in the different Fourier modes of the azimuthal temperature and azimuthal vertical velocity profile, to determine whether a single LSC is present. 
In section  \ref{sec_relativeLSCstrength2} we will look into time resolved properties of the
LSC.
The general conclusions of this paper will be presented in section \ref{sec_conclusions}.

\section{Numerical method and procedure} \label{sec_numericalmethod}
The flow is solved by numerically integrating the three-dimensional unsteady Navier-Stokes equations within the Boussinesq approximation. For a detailed discussion of the numerical code we refer to Refs. \cite{ver96,ver99,ver03,ste10}. The flow is simulated in a cylindrical sample in order to keep the geometry identical to the one used in most experiments. In \cite{zho09b,ste09,ste10,wei10} we have shown that simulation results obtained with this code agree excellent with experimental results. The cases we studied are based on the most common experimental setups that are available, namely $Pr\approx 6.4$ and $Pr\approx 0.7$ at an aspect ratio $\Gamma=D/L$ of $1$ and $1/2$. A detailed overview of the simulations can be found in table \ref{table1}. 

In the simulations we placed up to $64$ azimuthally equally spaced numerical probes per horizontal plane that provide simultaneous point-wise measurements of the temperature and the three velocity and vorticity components at the heights $0.25L$, $0.50L$, and $0.75L$ and a distance $0.45 D$ from the cylinder axis. Grid refinement tests were performed because the region close to the sidewall, where the LSC properties are sampled, is most sensitive from a resolution point of view \cite{ste10}. The simulations are fully resolved according to the criteria of Stevens et al. \cite{ste10} and the LSC properties we find do not depend on the grid resolution. Note that the azimuthal and radial number of grid points required to get the same resolution with respect to the turbulent length scales is less for the $\Gamma=1/2$ cases than for the $\Gamma=1$ cases because for the latter a larger horizontal extent has to be simulated. 

\begin{table}
  \centering
  \caption{The columns from left to right indicate the following: $Ra$, $Pr$, $\Gamma$, the number of grid points in the azimuthal, radial and axial directions ($N_{\theta} \times N_r \times N_z$). The last two columns indicate the relative LSC strength determined using equation (\ref{Eq Relative Strength LSC}), see section \ref{sec_relativeLSCstrength}, using the data of only $8$ ($\bar{S}_m(8)$) and all $64$ ($\bar{S}_m(64)$) probes in the horizontal midplane, respectively.}
  \label{table1}
\begin{tabular}{|c|c|c|c|c|c|}
  \hline
  $Ra$ &
  $Pr $ &
  $\Gamma$ &
  $N_{\theta} \times N_r \times N_z$ &
  $\bar{S}_m(8)$ &
  $\bar{S}_m(64)$
  \\
  \hline
  $1 \times 10^8$      &  $6.4$    & $1.0$  & $257    \times 129    \times 257$ &0.74 & 0.70 \\
  $1 \times 10^8$      &  $6.4$    & $1.0$  & $385    \times 193    \times 385$ &0.65 & 0.68 \\
  $5 \times 10^8$      &  $6.4$    & $1.0$  & $257    \times 129    \times 257$ & 0.68 & 0.73\\
  $5 \times 10^8$      &  $6.4$    & $1.0$  & $385    \times 193    \times 385$  &0.70 & 0.76 \\
  $1 \times 10^8$      &  $0.7$    & $1/2$  & $193    \times   65    \times 257$  &0.45 & 0.57\\  
  $1 \times 10^8$      &  $0.7$    & $1/2$  & $257    \times   97    \times 385$  &0.42 & 0.54\\    
  $1 \times 10^8$      &  $6.4$    & $1/2$  & $193    \times   65    \times 257$  &0.06 & 0.27\\ 
    \hline
\end{tabular}
\end{table}
 
\section{Methods to determine the LSC orientation} \label{sec_LSCmethod}

\begin{figure}
  \centering
  \subfigure{\includegraphics[width=3.25in]{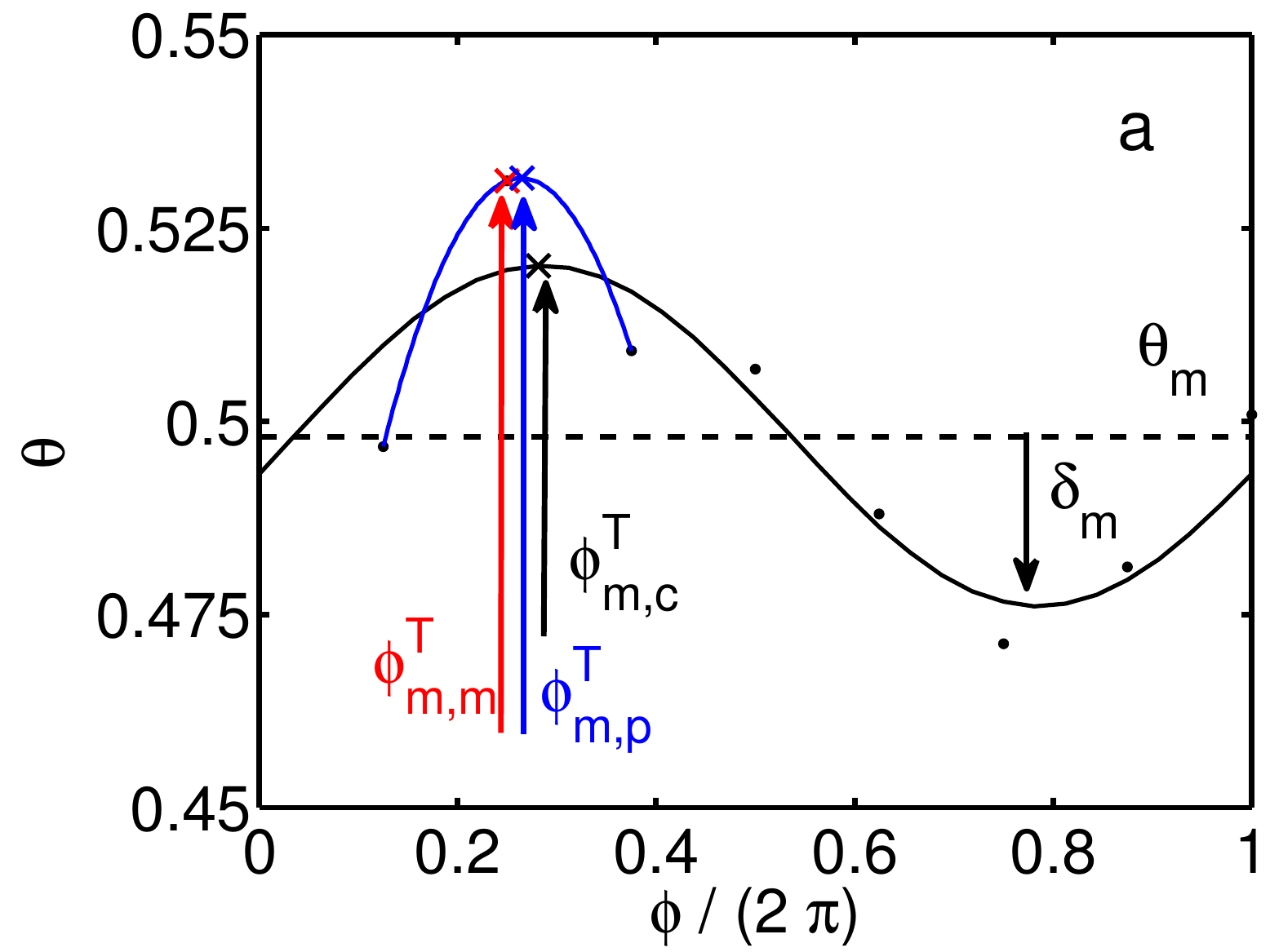}}
  \subfigure{\includegraphics[width=3.25in]{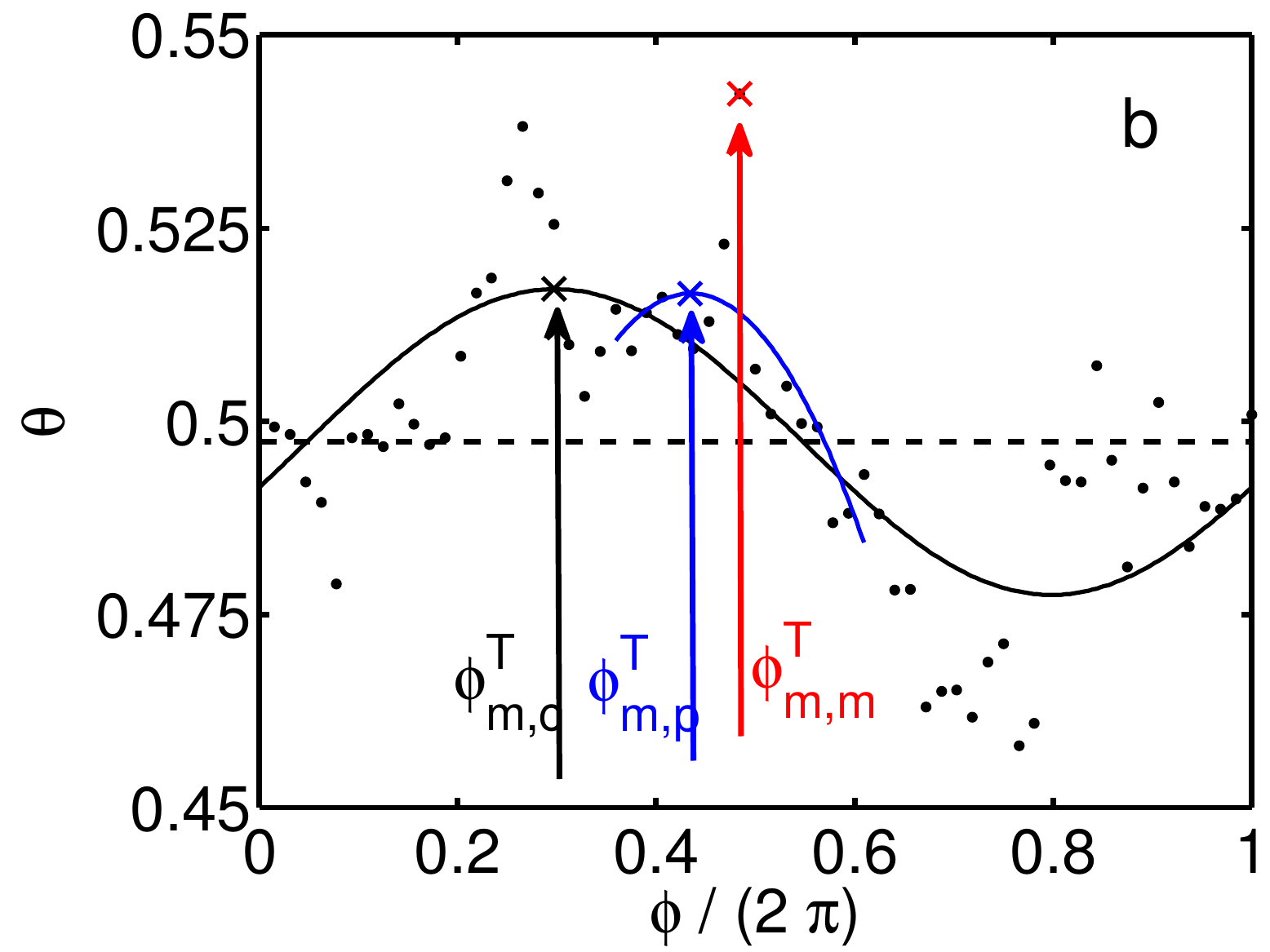}}
  \caption{Azimuthal temperature profile averaged over $4 \tau_f$ at midheight for $Ra=5\times10^8$, $Pr=6.4$, and $\Gamma=1$.
   The measured data from the numerical probes are indicated by the dots.
  The solid black line shows the cosine fit (eq. (\ref{Eq cosine fit})) to the data with offset $\theta_m$ (dashed line) and amplitude $\delta_m$. The black cross indicates the position of $\phi_{m,c}^T$. The red cross indicates the position of the probe that records the highest temperature and thus the position of $\phi_{m,m}^T$. The blue solid line gives the polynomial fit around the maximum. Its maximum, indicated by the blue cross, indicates the position of $\phi_{m,p}^T$. The $\phi(x)$ range that is considered in the polynomial fit is $\frac{1}{2}\pi$. Panel a and b shows the procedure applied to the data of $8$ and $64$ equally spaced probes at midheight, respectively. The corresponding azimuthal vertical velocity profiles are given in figure \ref{Fig_S10_Snapshotmovie}.}
  \label{Fig_S09b_Method}
\end{figure}

Following Ahlers and coworkers \cite{bro06,bro07,ahl09}, 
the orientation and strength of the LSC can be determined by fitting the function
\begin{equation}\label{Eq cosine fit}
    \theta_i = \theta_m + \delta_m \cos(\phi_i-\phi_m)
\end{equation}
to the temperatures recorded by the numerical probes at the height $z=0.5L$ and the azimuthal positions $\phi_i=2 i\pi/n_p$, where $n_p$ indicates the number of probes. The three parameters $\theta_m$, $\delta_m$, and $\phi_m$ are obtained from least square fits. Here $\delta_m$ is a measure of the temperature amplitude of the LSC and $\phi_m$ is the azimuthal orientation of the LSC at midheight. The azimuthal average of the temperature at the horizontal mid-plane is given by $\theta_m$. We calculated temperatures $\theta_t$ and $\theta_b$, orientations $\phi_t$ and $\phi_b$, and amplitudes $\delta_t$ and $\delta_b$ for the top and bottom levels at $z=0.75L$ and $z=0.25L$ separately by the same method. For experiments \cite{bro06,bro07,ahl09}
exactly this method is applied to the data of $8$ equally spaced thermistors, see figure \ref{Fig_S09b_Method}.

A second method that is applied to determine the LSC orientation is to make a polynomial fit around the sensor that records the highest (lowest) temperature \cite{xi09,zho09}. Finally, since we have a large number of numerical probes, i.e. up to $64$ azimuthally equally spaced per horizontal level, we also determine the LSC orientation by finding the probe that records the highest (lowest) temperature. Again the LSC orientation based on these two methods is determined separately for all three levels, see figure \ref{Fig_S09b_Method}.

To distinguish the different methods we introduce a notation with a second index, i.e. $\phi_{*,[c,p,m]}$. Here $c$ indicates that $\phi$ is determined using a cosine fit to the azimuthal temperature profile, $p$ that $\phi$ is determined using a polynomial fit around the maximum (or minimum) observed in the azimuthal profile, and $m$ indicates that the maximum (or minimum) observed in the azimuthal profile is used. Because the numerical probes record both the temperature and vertical velocity component we use both to determine the LSC orientation. In the rest of the paper we will use the notation $\phi_{*,*}^T$ to indicate that the orientation is based on the azimuthal temperature profile and $\phi_{*,*}^{w}$ when it is based on the azimuthal vertical velocity profile. Hence $\phi$ has three indices, i.e. $\phi_{1,2}^3$, see table \ref{table2} for a detailed overview. In the rest of the paper we will indicate the numerical value of an index when this index is varied.

Because the numerically obtained azimuthal profiles can be very noisy, we applied a moving averaging filter to the data obtained from the numerical probes to eliminate the detection of very small plumes. In experiments these events are not detected anyhow by thermistors, since these need time to react to temperature changes in the flow. We decided to apply a moving averaging filter of $4 \tau_f$, where $\tau_f$ is defined with respect to the free-fall velocity $U_f$ as $\tau_f = L/U_f$ ($L$ is the height of the sample). Recently, Bailon-Cuba et al. \cite{bai10} showed the characteristic convective velocity $U_c$ of the LSC is approximately $U_f/5$ for the parameter ranges $Ra=10^7-10^9$, $\Gamma=1/2-12$, and $Pr=0.7$. This means that the LSC turnover time $\tau_{LSC}$ defined as $\tau_{LSC}=2L/U_c$ is about 10 $\tau_f$.

\begin{table}
  \centering
  \caption{The LSC orientation in this paper is determined from the azimuthal temperature or azimuthal vertical velocity profile with different methods. In addition, the LSC orientation is also determined at different heights. As explained in the text this information is indicated in the notation $\phi_{1,2}^3$. In this table the meaning of the symbols at the index locations $1$, $2$, and $3$, is summarized.}
  \label{table2}
\begin{tabular}{|c|c|c|}
  \hline
  Index &
  Symbol  &
  Meaning \\
  \hline
  $1$      &  $b$        & Height $0.25L$  \\
  $1$      &  $m$       & Height $0.50L$   \\
  $1$      &  $t$         & Height $0.75L$  \\
  $2$      &  $c$        & Cosine fit  \\
  $2$      &  $p$        & Polynomial fit   \\  
  $2$      &  $m$       & Maximum   \\    
  $3$      &  $T$        & Azimuthal temperature profile   \\    
  $3$      &  $w$       & Azimuthal vertical velocity profile   \\    
    \hline
\end{tabular}
\end{table}

\section{Results for $\Gamma=1$} \label{sec_Gamma1}

\begin{figure}
  \centering
  \subfigure{\includegraphics[width=3.25in]{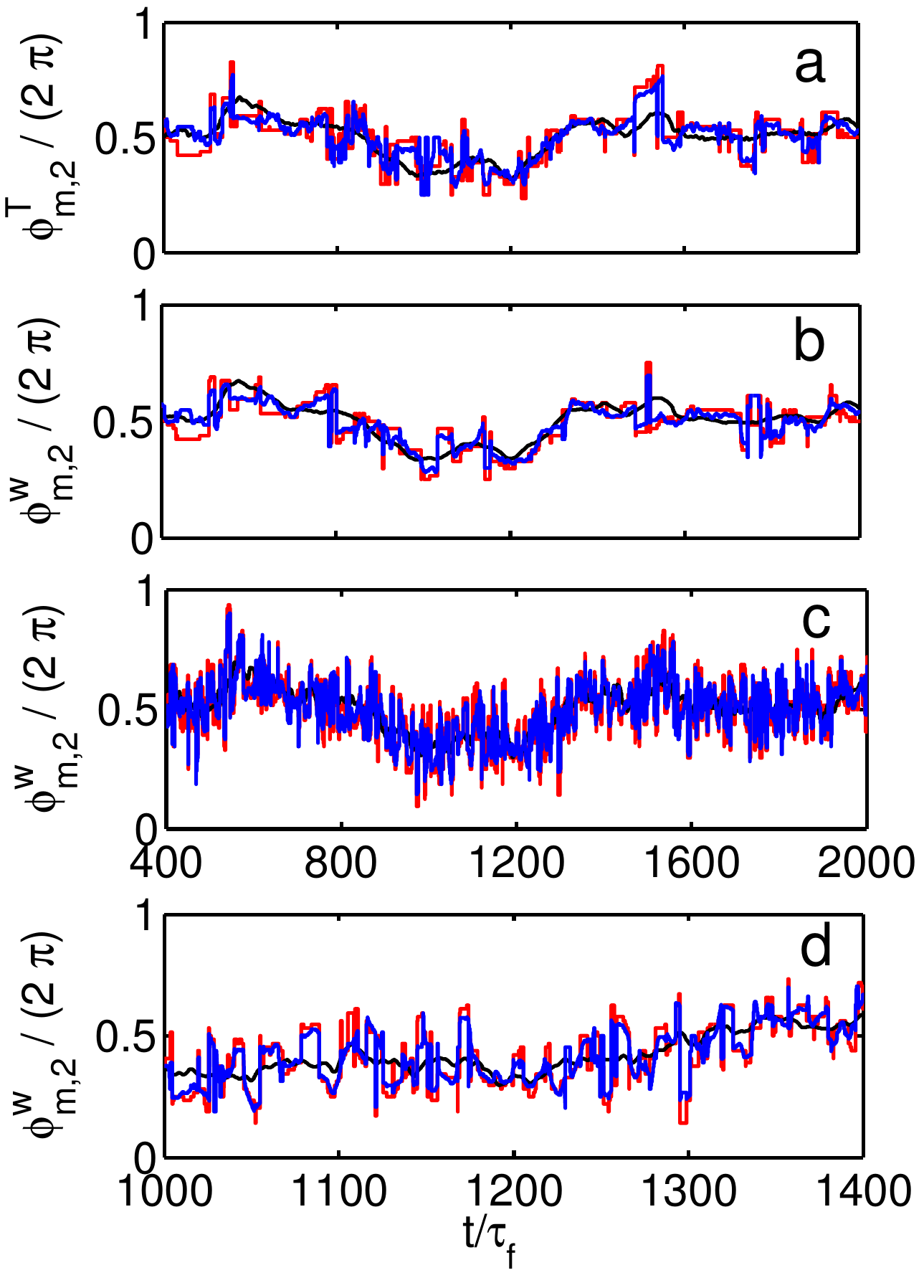}}
  \caption{The LSC orientation for $Ra=1\times10^8$, $Pr=6.4$, and $\Gamma=1$ on the $257\times129\times257$ grid based on the data of $64$ azimuthally equally spaced probes per horizontal level. The black, red, and blue line indicate $\phi_{m,c}^3$ $\phi_{m,m}^3$, and $\phi_{m,p}^3$, respectively. a) $\phi_{m,2}^T$ based on the azimuthal temperature profile averaged over $50 \tau_f$ ($\approx 5$ LSC turnover times), b) $\phi_{m,2}^{w}$ based on the azimuthal vertical velocity profile averaged over $50 \tau_f$, c) $\phi_{m,2}^{w}$ based on the azimuthal vertical velocity profile averaged over $4 \tau_f$. d) Enlargement of the graph shown in panel c to reveal more details.}
  \label{Fig_S09b_LSCposition}
\end{figure}

\begin{figure}
  \centering
  \subfigure{\includegraphics[width=3.25in]{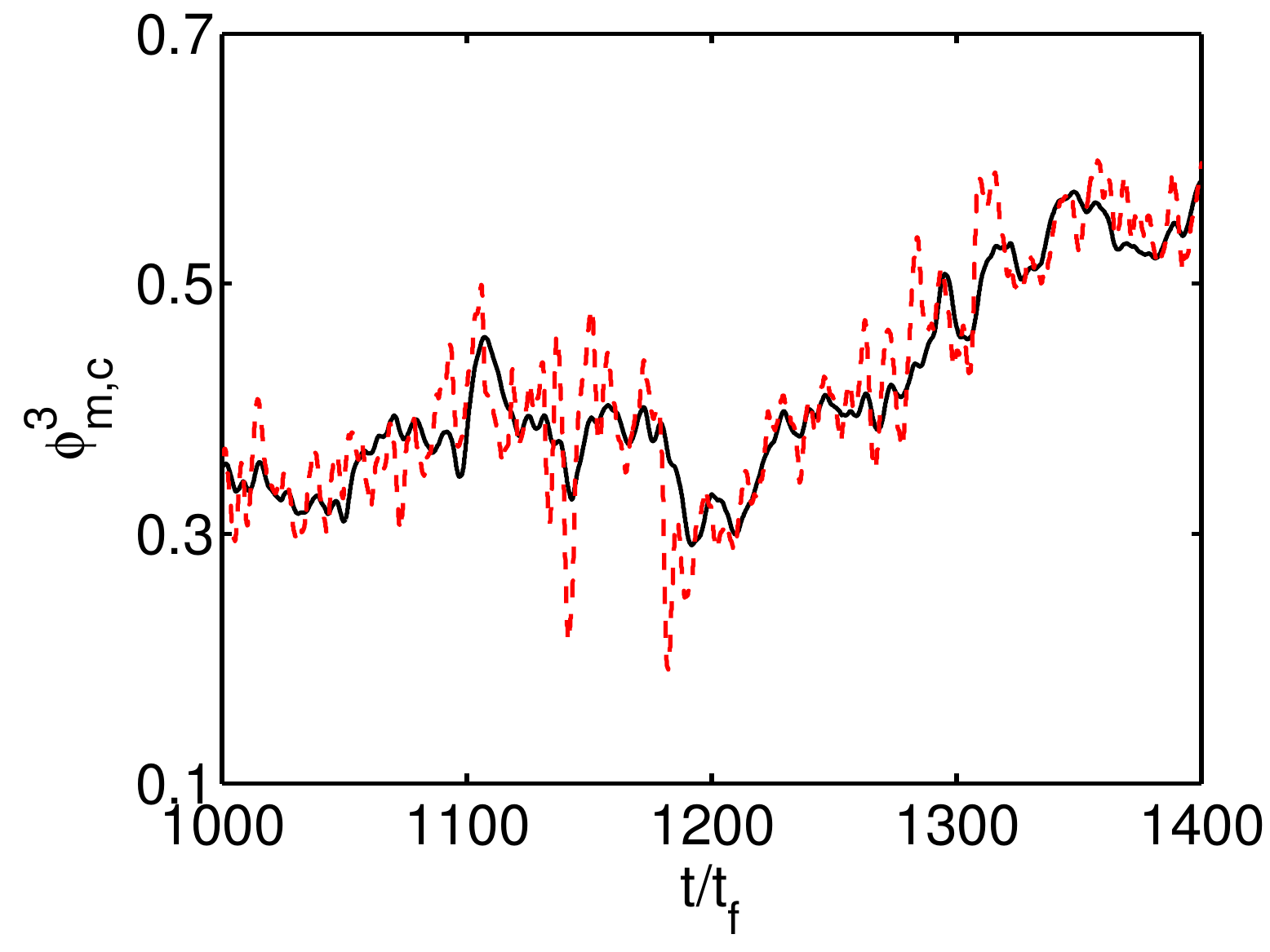}}
  \caption{$\phi_{m,c}^{w}$ (solid black line) and $\phi_{m,c}^{T}$ (red dashed line) for $Ra=1\times10^8$, $Pr=6.4$, and $\Gamma=1$ on the $257\times129\times257$ grid, and based on the azimuthal vertical velocity and azimuthal temperature profile averaged over $4 \tau_f$. Here the data of all $64$ azimuthally equally spaced probes at midheight have been used.}
  \label{Fig_S10_LSC_tempvsW}
\end{figure}

Figure \ref{Fig_S09b_LSCposition} shows the typical behavior of the LSC orientation for $Pr=6.4$ in a $\Gamma=1$ sample. The figure shows $\phi_{m,2}^T$ and $\phi_{m,2}^{w}$. In order to show that the LSC orientation based on the temperature and the vertical velocity data is almost identical we consider first a long averaging time ($50 \tau_f$). The results are displayed in figure \ref{Fig_S09b_LSCposition}a and b. The similarity is expected since the LSC carries warm fluid from the bottom plate up the sidewall and vice versa. To show the influence of this long time averaging we show $\phi_{m,2}^{w}$ based on the azimuthal vertical velocity profile after averaging over $4 \tau_f$, see figure \ref{Fig_S09b_LSCposition}c and \ref{Fig_S09b_LSCposition}d. When no time averaging is applied the graph looks similar, with some additional peaks due to very small plumes, to the one where the data is averaged over $4 \tau_f$. Figure  \ref{Fig_S10_LSC_tempvsW} shows that the LSC orientation can be determined more precisely from the vertical velocity than from the temperature, even for this relatively high $Pr$ (i.e. small thermal diffusivity). The difference between the LSC orientation determined from the vertical velocity and the temperature seems to depend on several parameters such as the $Ra$ and the $Pr$ number, and the aspect ratio. However, at the moment we do not have enough numerical data available to systematically study this. For $\Gamma=1$ the obtained LSC orientation at midheight is the same when the data of only $8$ or all $64$ probes is used, but not when the data of only $2$ probes is used.

\begin{figure}
  \centering
  \includegraphics[width=3.25in]{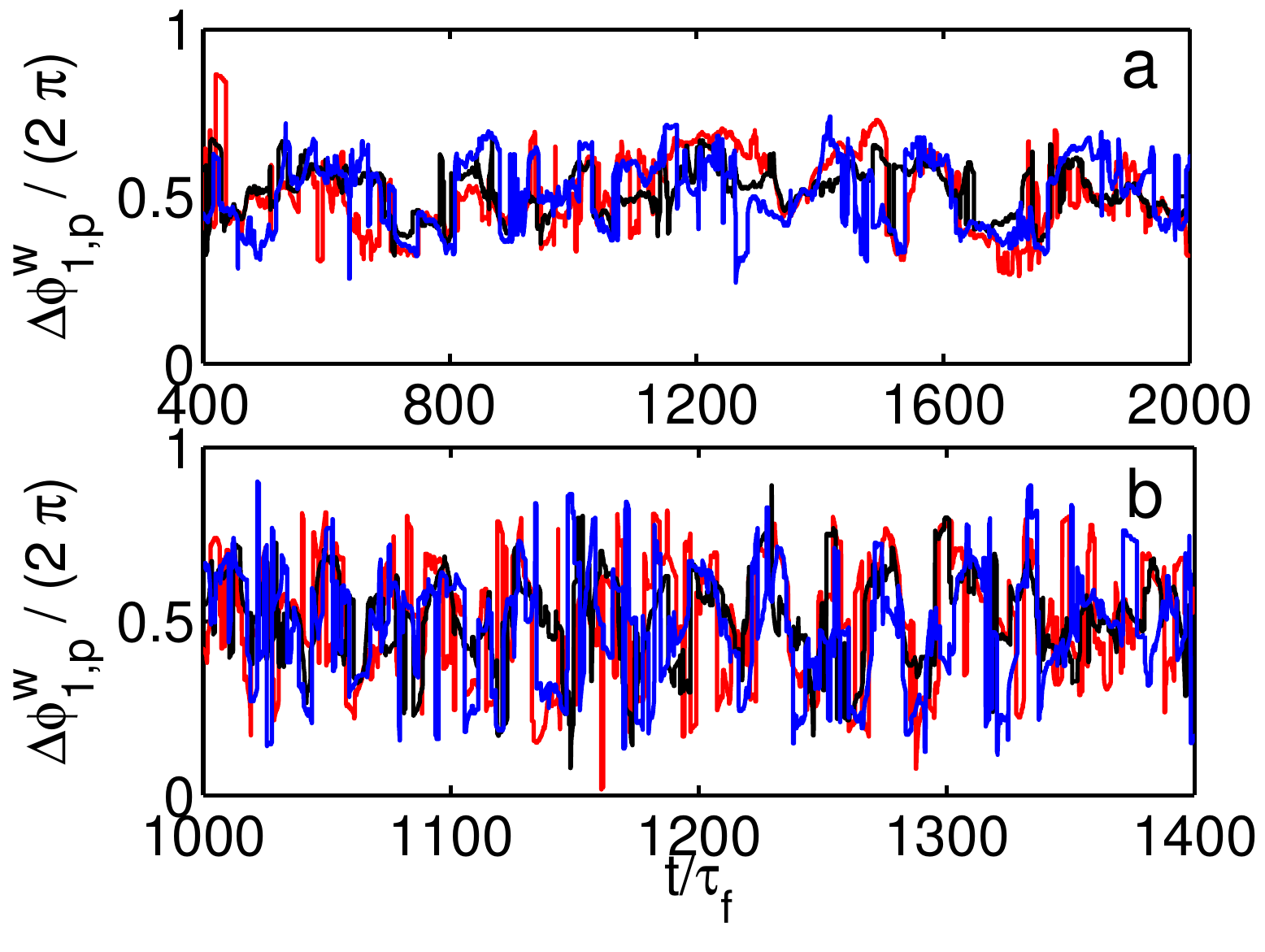}
  \caption{The red, black, and blue lines indicate $\Delta\phi^{w}_{b,p}$, $\Delta\phi^{w}_{m,p}$, and $\Delta\phi^{w}_{t,p}$, respectively for $Ra=1\times10^8$, $Pr=6.4$, and $\Gamma=1$ on the $257\times129\times257$ grid. Panel a and b show $\Delta\phi^{w}_{1,p}$ when the azimuthal vertical velocity profile is averaged over $50 \tau_f$ and $4 \tau_f$, respectively. The average of $\Delta\phi^{w}_{1,p}$ is $\pi$ and this is in agreement with a cosine fit. The deviations from this value are due to plumes and the off-center motion of the LSC. Here the data of all $64$ azimuthally equally spaced probes per horizontal level have been used.}
  \label{Fig_S09b_offcenter}
\end{figure}

\begin{figure}
  \centering
  \includegraphics[width=3.25in]{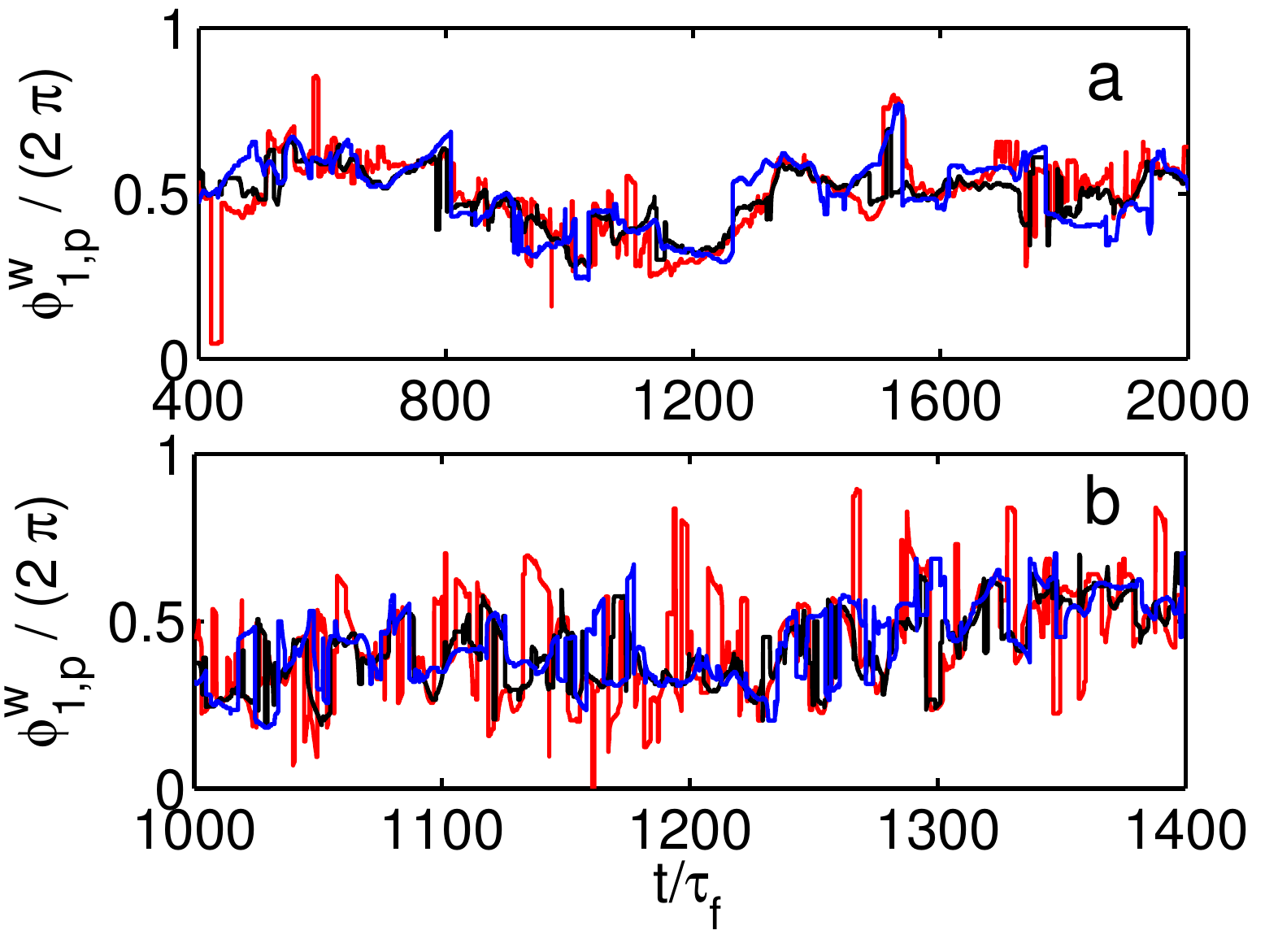}
  \caption{The red, black, and blue lines indicate $\phi_{b,p}^{w}$, $\phi_{m,p}^{w}$, and $\phi_{t,p}^{w}$, respectively, for $Ra=1\times10^8$, $Pr=6.4$, and $\Gamma=1$ on the $257\times129\times257$ grid. Panel a and b show $\phi^{w}_{1,p}$ based on the azimuthal profiles of the vertical velocity averaged over $50 \tau_f$ and $4 \tau_f$, respectively. The difference between the LSC orientation at the different levels is due to plumes and the torsional motion of the LSC. Here the data of all $64$ azimuthally equally spaced probes per horizontal level have been used.}
  \label{Fig_S09b_torsial}
\end{figure}

From figure \ref{Fig_S09b_LSCposition} it becomes clear that the LSC orientation, obtained by the polynomial and cosine fit, can differ significantly. This is due to the off-center motion of the LSC \cite{xi09,zho09,bro09}. Figure \ref{Fig_S09b_offcenter} shows $\Delta\phi=\phi_{max}-\phi_{min}$, i.e. the difference between the orientation of the strongest up ($\phi_{max}$) and down going motion ($\phi_{min}$) obtained using the polynomial fit, fluctuates around $\Delta\phi=\pi$. To quantify the strength of these fluctuations we calculate $\langle(\Delta\phi^w_{1,p}-\pi)^2\rangle^{1/2}$. The values based on the data of $64$ ($8$) probes per horizontal level for the case presented in figure \ref{Fig_S09b_offcenter} are $\langle(\Delta\phi^w_{t,p}-\pi)^2\rangle^{1/2}\approx 1.09 (1.04)$, $\langle(\Delta\phi^w_{b,p}-\pi)^2\rangle^{1/2} \approx 0.84 (0.77)$, and $\langle(\Delta\phi^w_{m,p}-\pi)^2\rangle^{1/2} \approx 1.11 (1.01)$. Because the fluctuation strength is the same when the data of $8$ and $64$ equally spaced probes per horizontal level are considered we conclude that the use of $8$ probes is sufficient to capture these statistics in a $\Gamma=1$ sample. In figure \ref{Fig_S09b_torsial} it is shown that the LSC orientation, obtained using the polynomial fit method, can be different at the different heights. We note that the same is observed when the LSC orientation based on a cosine fit is considered. This indicates that the LSC is not always flowing straight up and down, but is also moving in the azimuthal direction, and thus performs twisting motions \cite{fun04}. We note that some phenomena, like the drift of the LSC due to the Coriolis force, see Ref.\ \cite{bro06b}, can be analyzed using the LSC orientation based on the polynomial and cosine method, because in this case only the long term drift of the LSC is important.

\begin{figure}
  \centering
  \includegraphics[width=2.0in]{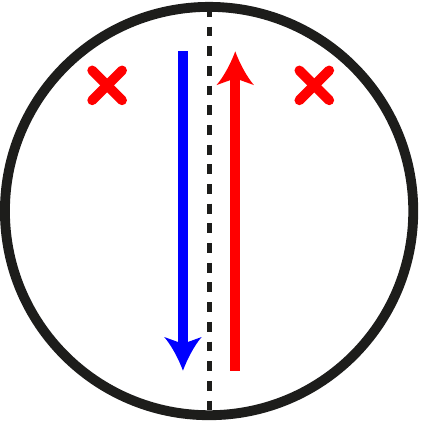}
  \caption{Sketch of the top view on the RB sample. The dotted line indicates the orientation of the LSC based on the cosine fit and the red (up-pointing) and blue (down-pointing) arrows indicated the flow direction close to the bottom and top plate, respectively. Strong plumes are detected alternatingly on the left and right side of the LSC, see crosses, with a specific frequency, see also figure \ref{Fig_S09b_LSCposition}. }
  \label{fig:sketch}
\end{figure}

\begin{figure}
  \centering
  \includegraphics[width=3.25in]{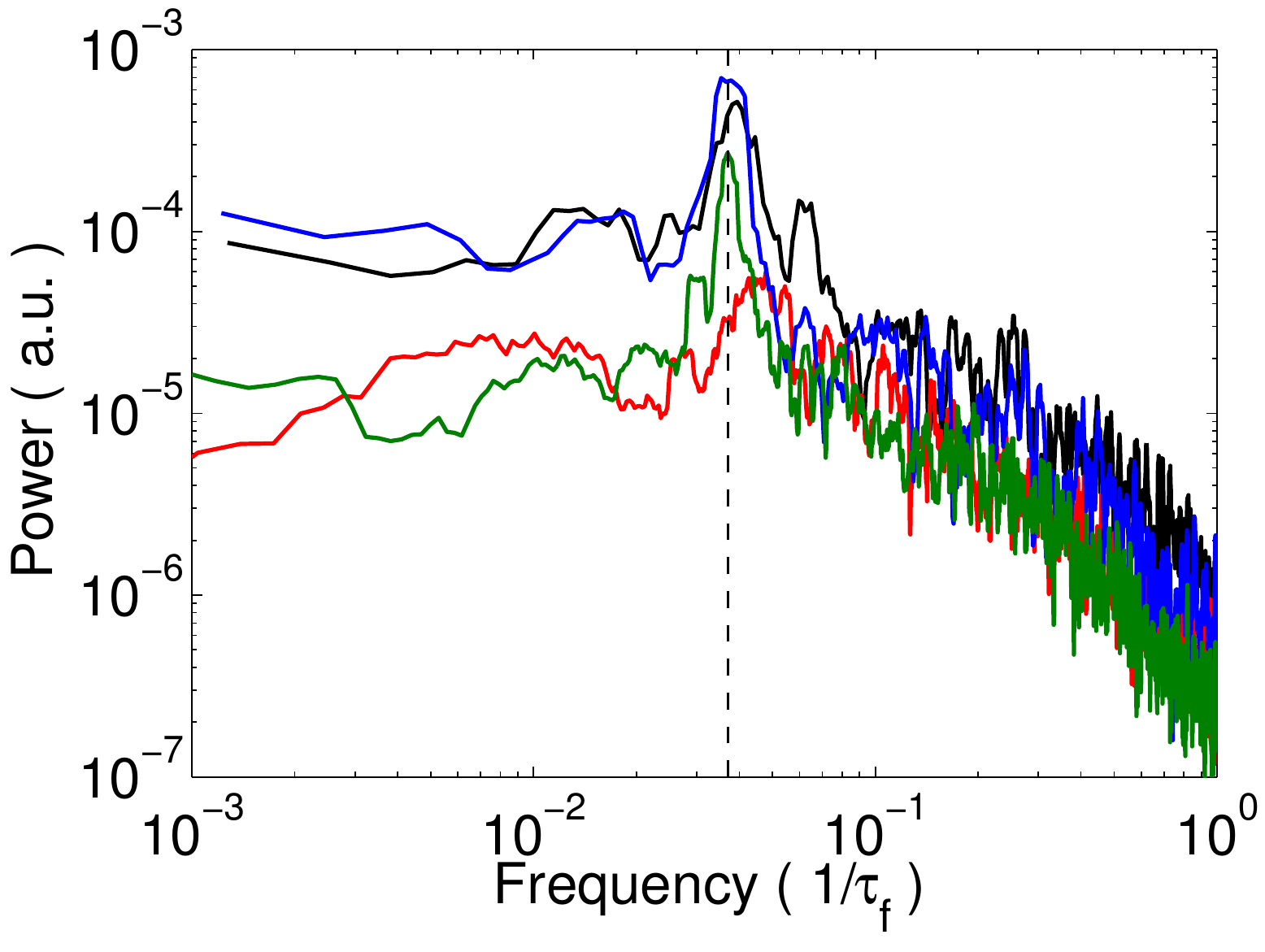}
  \caption{The power spectra of $\phi_{m,c}^{w} - \phi_{m,p}^{w}$ shows a maximum around $25 \tau_f$, which is indicated by the vertical dashed line. This frequency is caused by plumes that pass the LSC orientation based on the cosine fit on the left and right side . The different spectra are obtained from simulations for $Pr=6.4$ in a $\Gamma=1$ sample for different $Ra$ and resolution, i.e. $Ra=1\times10^8$ on a  $257 \times 129 \times 257$ (red) and on a $385 \times 193 \times 385$ grid (black), and $Ra=5\times10^8$ on a $257 \times 129 \times 257$ (green) and a $385 \times 193 \times 385$ (blue) grid.}
  \label{Fig_S09b_Spectra}
\end{figure}

\begin{figure}
  \centering
  \subfigure{\includegraphics[width=3.25in]{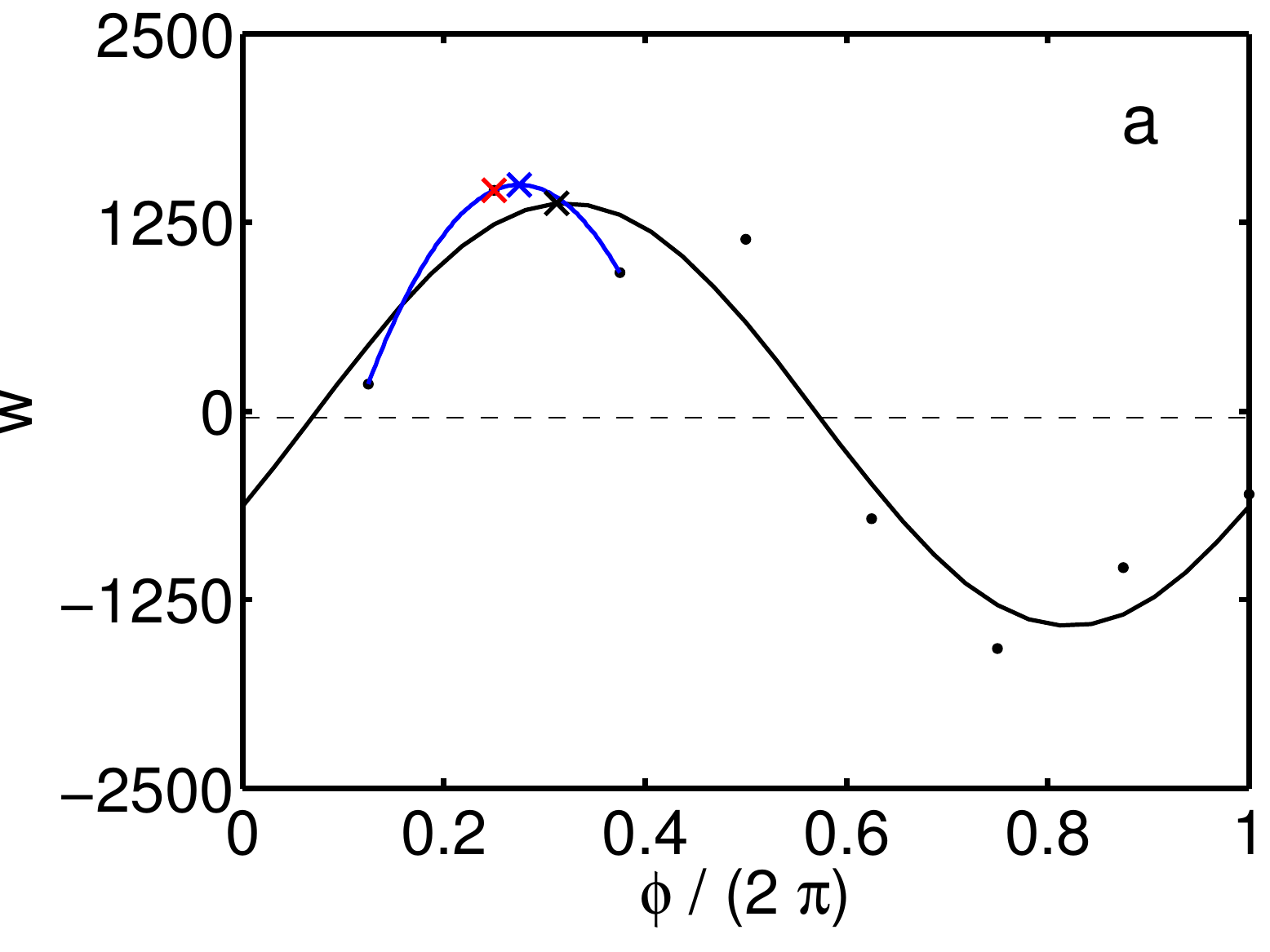}}
  \subfigure{\includegraphics[width=3.25in]{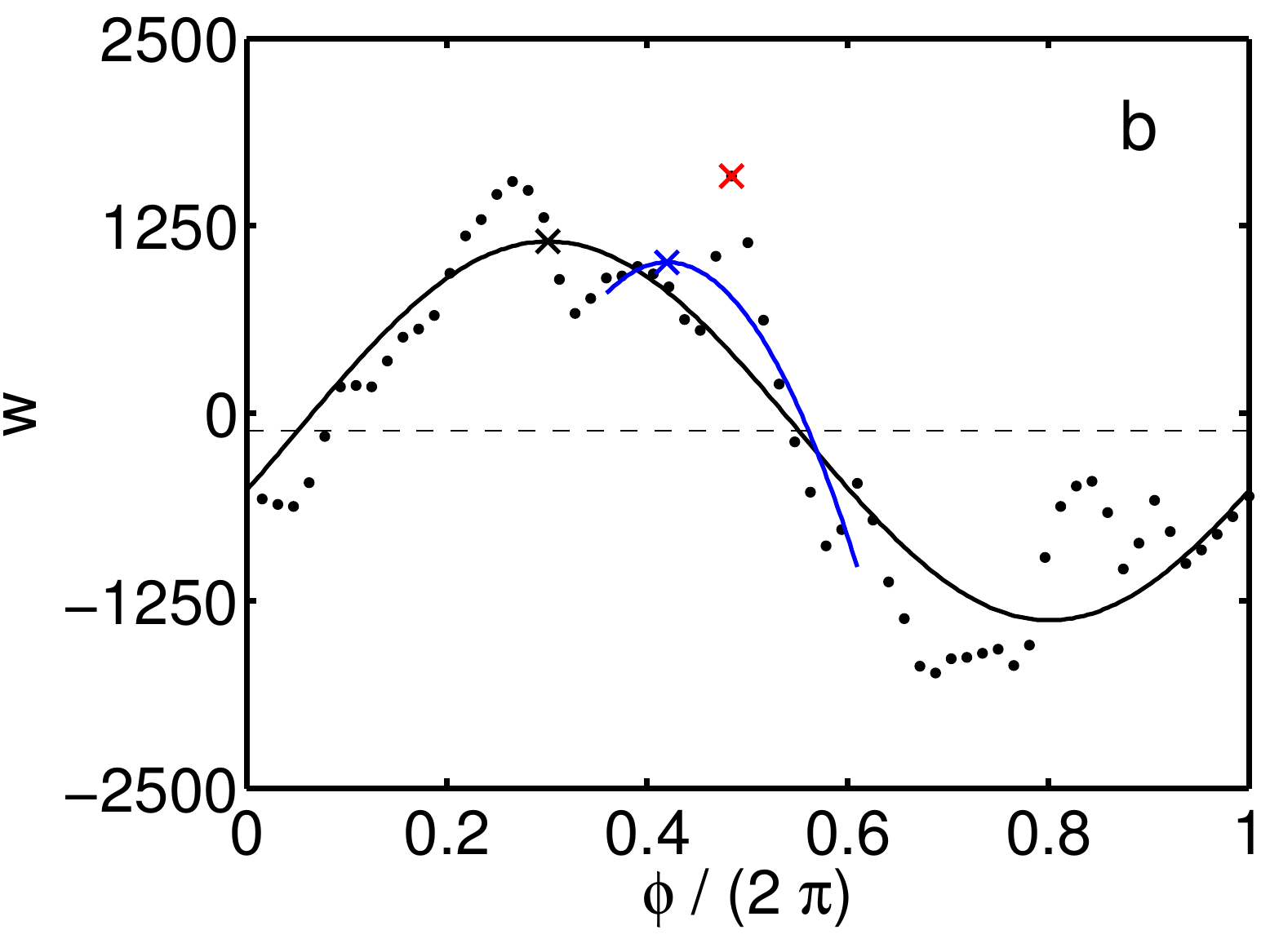}}
  \caption{Snapshot of the azimuthal vertical velocity profile averaged over $4 \tau_f$ for $Pr=6.4$ in a $\Gamma=1$ sample at $Ra=5\times10^8$. The symbols have the same meaning as indicated in figure \ref{Fig_S09b_Method}.  Panel a indicates the obtained profile using the data of $8$ probes and panel b the profile as it is obtained from the data of $64$ probes. The accompanying movies can be found in the supplementary material \cite{ste10movies}.}
  \label{Fig_S10_Snapshotmovie}
\end{figure}

More remarkably is the presence of a specific frequency of approximately $25 \tau_f$ in the signal $\phi_{m,c}^w - \phi_{m,p}^w$, see figure \ref{Fig_S09b_Spectra}. This specific frequency seems to be related to the frequency in which plumes are passing the horizontal midplain \cite{vil95}, which in our case shows up in the frequency in which plumes are detected on the left and right hand side of the LSC orientation based on the cosine fit, see figure \ref{fig:sketch}. This frequency might also be related to the low frequency mode found already in Ref. \cite{cas89} or to the off-center oscillation of the LSC \cite{zho09b}.
 We find this frequency in our simulations for $Pr=6.4$ in a $\Gamma=1$ sample. We note that we also find this frequency when we do not apply any time averaging on the data before we determine the LSC orientation. Actually one can already see this phenomenon in figure \ref{Fig_S09b_LSCposition}d where $\phi_{m,p}^{w}$ fluctuates around $\phi_{m,c}^{w}$ with a typical period of about $25 \tau_f$. Further confirmation is obtained when the temporal behavior of the data obtained from the $64$ probes at midheight is shown in a movie, see the supplementary material \cite{ste10movies}. The movie is based on the data obtained from the simulation on the $385 \times 193 \times 385$ grid at $Ra=5\times 10^8$ with $Pr=6.4$ in a $\Gamma=1$ sample. The movie shows a passing plume at t $\approx 2412$, see the snapshot of the movie in figure \ref{Fig_S10_Snapshotmovie}b, which is indicated by the two peak structure around the maximum position obtained by the cosine fit. When this plume is passed the double peak structure disappears. Subsequently, the LSC orientation based on the polynomial fit method is smoothly passing the LSC orientation based on the cosine fit. Thus showing an off-center motion of the LSC. We note that having a large number of probes is essential to confirm these plume events. To show this we made a movie of the same time interval, see the supplementary material \cite{ste10movies} and the snapshot in figure \ref{Fig_S10_Snapshotmovie}a, with the data of only $8$ probes. From these movies and figure \ref{Fig_S10_Snapshotmovie}a we conclude that it is impossible to see the double peak structure when the data of only $8$ probes is used and therefore we cannot distinguish the off-center motion of the LSC from passing plumes, when the plumes stretches over two rows of thermistors.

In summary, in a $\Gamma=1$ sample the obtained LSC orientation using a cosine and polynomial fit is the same when the data of only $8$ or all $64$ azimuthally equally spaced probes per horizontal level are used. Because the azimuthal temperature profile is only an indirect measure of the LSC orientation we find that the LSC orientation can be determined with more precision from the azimuthal vertical velocity profile. In addition,
 we find that many of its interesting phenomena like azimuthal meandering, reversals, cessations, and the sloshing mode of the LSC look the same when the data of only $8$ or all $64$ azimuthally equally spaced probes per horizontal level are considered. However, only with the data of $64$ equally spaced probes per horizontal level the effect of passing plumes on the azimuthal temperature and azimuthal vertical velocity profiles can be identified.

\section{Results for $\Gamma=1/2$} \label{sec_Gamma05}

We now come to the $\Gamma =1/2$ case. For this geometry Verzicco and Camussi \cite{ver03} have used the data accessibility provided by DNS to show that the large scale circulation can be either in a single-roll state (SRS) or in a double-roll state (DRS). Subsequently, a model introduced by Stringano and Verzicco \cite{str06} found that the switching between the DRS and the SRS can influence the Nusselt number, and can thus be a possible reason for the bimodal behavior of Nusselt found in some experiments, see e.g.\ \cite{roc02,roc10}.

The first extensive experimental study on the transition between the SRS and the DRS was performed by Xi and Xia \cite{xi08}. They studied flow mode transition in samples of aspect ratio $1$, $1/2$ and $1/3$. Figure 10 of their paper \cite{xi08} shows the percentages of the 
time the flow spends in either the SRS or the DRS. In addition, they showed that the heat transfer in the SRS is higher than in the DRS. Recently, Weiss \& Ahlers \cite{wei10b} have also experimentally investigated the switching between the SRS and the DRS, but now over a much larger range of $Ra$ than Xi and Xia \cite{xi08}. The  conclusion of this work is summarized in figure 11 of Ref.\ 
 \cite{wei10b}. That figure shows that the SRS is more dominant at higher $Ra$ and smaller $Pr$. They note that they find a good agreement with the results of Xi and Xia in the $Ra$ number range in which both experiments overlap. 
In addition, the work of Xi and Xia \cite{xi08} and Weiss and Ahlers \cite{wei10b} showed that the flow state can not always 
be defined as SRS or DRS.  This is particularly important for $Ra \lesssim 3\times10^9$. 
E.g., 
for $Ra=1\times10^8$ and $Pr=4.38$ the flow state is undefined for about $50\%$ of the time according to Weiss and Ahlers \cite{wei10b}.

 \begin{figure}
  \centering
  \includegraphics[width=3.25in]{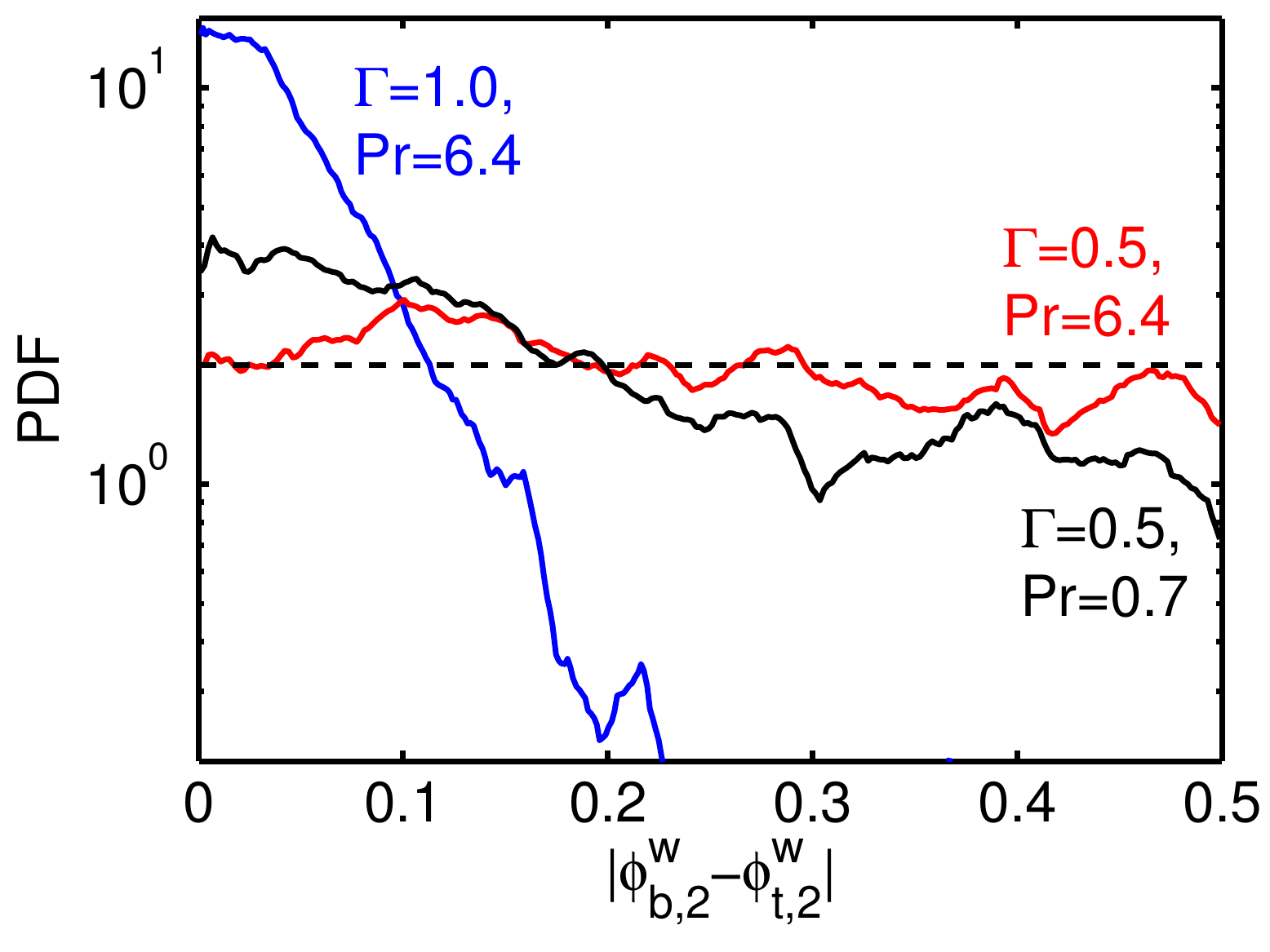}
  \caption{The PDF of $|\phi_{b,c}^w-\phi_{t,c}^w|$ for $Ra=1\times10^8$. The PDF for $Pr=6.4$ and $\Gamma=1$ is given in blue. The red and black line give the PDF for $Pr=6.4$ and $Pr=0.7$ in the $\Gamma=1/2$ geometry, respectively.}
  \label{Fig_S10_rollstate}
\end{figure}

\begin{figure}
  \centering
  \includegraphics[width=3.25in]{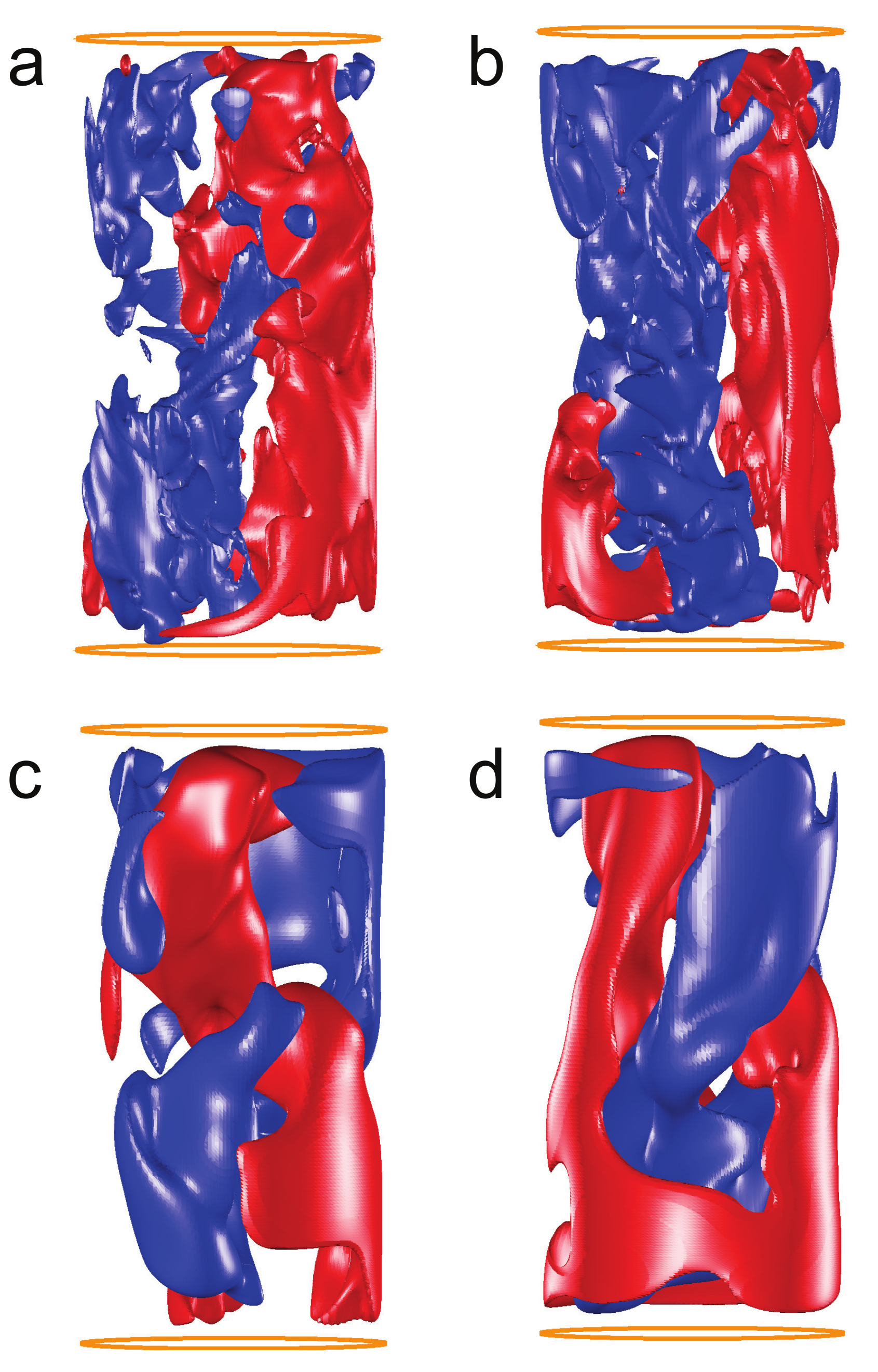}
  \caption{Snapshot of the regions with the strongest, i.e. $w \geqq 0.35 w_{max}$, up (red) and down-flow (blue) for $Ra=1\times10^8$ and $\Gamma=1/2$. a) $Pr=0.7$, SRS, b) $Pr=0.7$, very strong corner flows, c) $Pr=6.4$, DRS, d) $Pr=6.4$, strong plume/torsional motion. Supplementary movies that offer a full three dimensional view are provided in the supplementary material \cite{ste10movies}.}
  \label{Fig_S10_rollstate2}
\end{figure}

In this paper we will qualitatively confirm these experimental and numerical findings. We will restrict ourselves to
this relatively low $Ra$ number regime, namely to
$Ra=1\times10^8$:  First, in order to directly compare with
the $\Gamma = 1$ case of the previous section, second, because 
current high $Ra$ number simulations \cite{ste10,ste10d} are too limited in the number of turnover times to obtain sufficient statistics on the different flow states.

Our  simulations confirm Weiss \& Ahlers's \cite{wei10b} finding  that there is more
 disorder in a $\Gamma=1/2$ sample than in a $\Gamma=1$ sample. 
We find that in this low $Ra$ number regime it is important to have a large number of probes to accurately represent the azimuthal temperature and azimuthal vertical velocity profiles to analyze the flow. In this regime we see that deflection of plumes due to corner rolls and other plumes happens more often than in a $\Gamma=1$ sample. These effects can be observed in the supplementary movies for $\Gamma=1/2$ at $Ra=1\times10^8$ and $Pr=0.7$ and $Pr=6.4$ \cite{ste10movies}. In the movies that show horizontal cuts at the heights $0.25L$, $0.50L$, and $0.75L$ one can see that the hot and cold regions interchange more often in a $\Gamma=1/2$ sample than in a $\Gamma=1$ sample.
 
In order to determine whether there is a SRS or a DRS in this geometry we determined the probability density function (PDF) of the quantity $|\phi_{b,c}^w-\phi_{t,c}^w|$. When this PDF peaks around zero it means that the fluid is flowing straight up, and this indicates the presence of a SRS. When there is a DRS the PDF should peak around $\pi$. Figure \ref{Fig_S10_rollstate} shows that for $Pr=6.4$ in a $\Gamma=1$ sample the PDF peaks around zero and this confirms the dominance of the SRS in this geometry.  For $Pr=0.7$ in a $\Gamma=1/2$ sample the figure shows that there is a small peak around zero, but it is much less pronounced than for the $\Gamma=1$ case. For $Pr=6.4$ in a $\Gamma=1/2$ sample the figure shows that the PDF is nearly uniform, which would suggest that neither the SRS or the DRS is dominant. This agrees with the results of Weiss \& Ahlers 
 \cite{wei10b}, since they find that for $Pr=4.38$ and $Ra \lesssim 3\times 10^9$ the flow state is poorly defined and no state dominates. Figure \ref{Fig_S10_rollstate2} reveals the origin of these nearly uniform PDFs in the $\Gamma=1/2$ sample by showing three-dimensional visualizations of the regions where the vertical velocities are strongest. An analysis of these (and similar figures) revealed that the uniform PDF is due to the possibility of different flow states, i.e. the SRS (figure \ref{Fig_S10_rollstate2}a) with $|\phi_{b,c}^w-\phi_{t,c}^w| \approx 0$, the DRS (figure \ref{Fig_S10_rollstate2}c) with $|\phi_{b,c}^w-\phi_{t,c}^w| \approx \pi$, strong corner flows (figure \ref{Fig_S10_rollstate2}b) which can have any $|\phi_{b,c}^w-\phi_{t,c}^w|$, and strong plume and/or torsional motions (figure \ref{Fig_S10_rollstate2}d) which also 
can have any $|\phi_{b,c}^w-\phi_{t,c}^w|$.

 \begin{figure}
  \centering
  \includegraphics[width=3.25in]{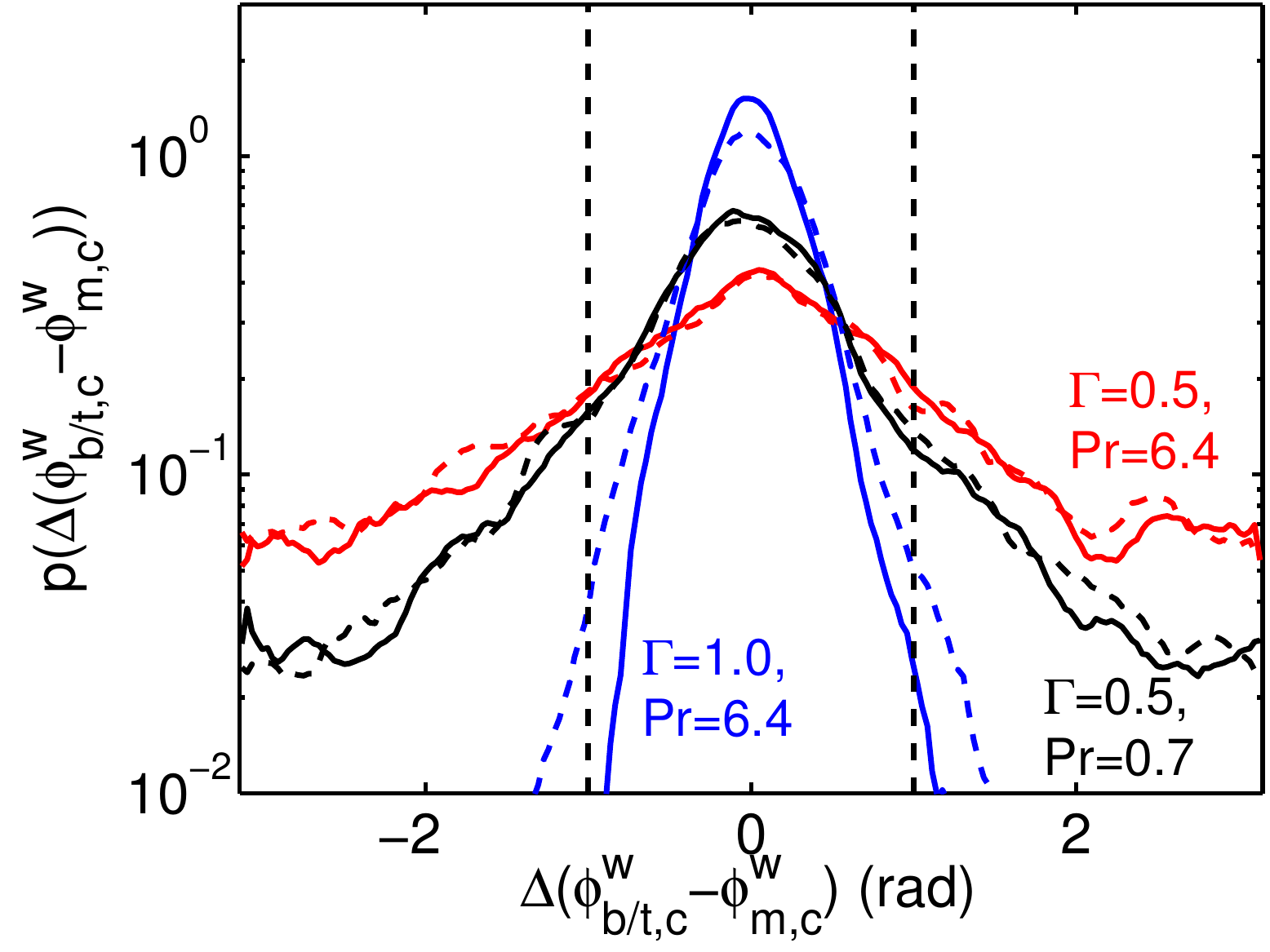}
  \caption{The PDF of $\phi_{b/t,c}^w-\phi_{m,c}^w$ for $Ra=1\times10^8$, see text for details. The dashed and solid lines indicate the PDF based on the data of only $8$ or all $64$ probes equally spaced probes per horizontal level, respectively. The vertical dashed lines indicate the definition of the SRS according to Xi and Xia \cite{xi08} and Weiss \& Ahlers  \cite{wei10b}.} 
  \label{Fig_S10_pdfGuenter}
\end{figure}

To better compare our results with the results of Weiss \& Ahlers 
 \cite{wei10b} we also determined the PDF of the quantity $\phi_{b/t,c}^w-\phi_{m,c}^w$, see figure \ref{Fig_S10_pdfGuenter}. Here $\phi_{b/t,c}^w$ means that both $\phi_{b,c}^w$ and $\phi_{t,c}^w$ are compared with $\phi_{m,c}^w$ to construct the PDF. A similar PDF for higher $Ra$, based on experimental data, can be found in figure 21 of Weiss \& Ahlers \cite{wei10b}. Figure \ref{Fig_S10_pdfGuenter} shows this PDF based on the data of only $8$ and all $64$ equally spaced probes per horizontal level for the
cases of figure \ref{Fig_S10_rollstate}. 
First of all,  
 figure  \ref{Fig_S10_pdfGuenter}
shows that the DRS is much more pronounced for $Pr=6.4$ than for $Pr=0.7$, which is in agreement with the results of Weiss and Ahlers \cite{wei10b} and the results shown in figure \ref{Fig_S10_rollstate}.
 Second,
 a comparison of the PDFs in figure \ref{Fig_S10_pdfGuenter} reveals that the PDFs based on the data of $8$ and all $64$ equally spaced probes are almost identical. The largest difference is visible for $Pr=6.4$ in a $\Gamma=1$ sample. Here the PDF based on the data of $64$ equally spaced probes is more confined in the SRS region than the one based on $8$ equally spaced probes, because the larger number of probes reduces 'random' fluctuations in the obtained LSC orientation.

\begin{figure}
  \centering
  \includegraphics[width=3.25in]{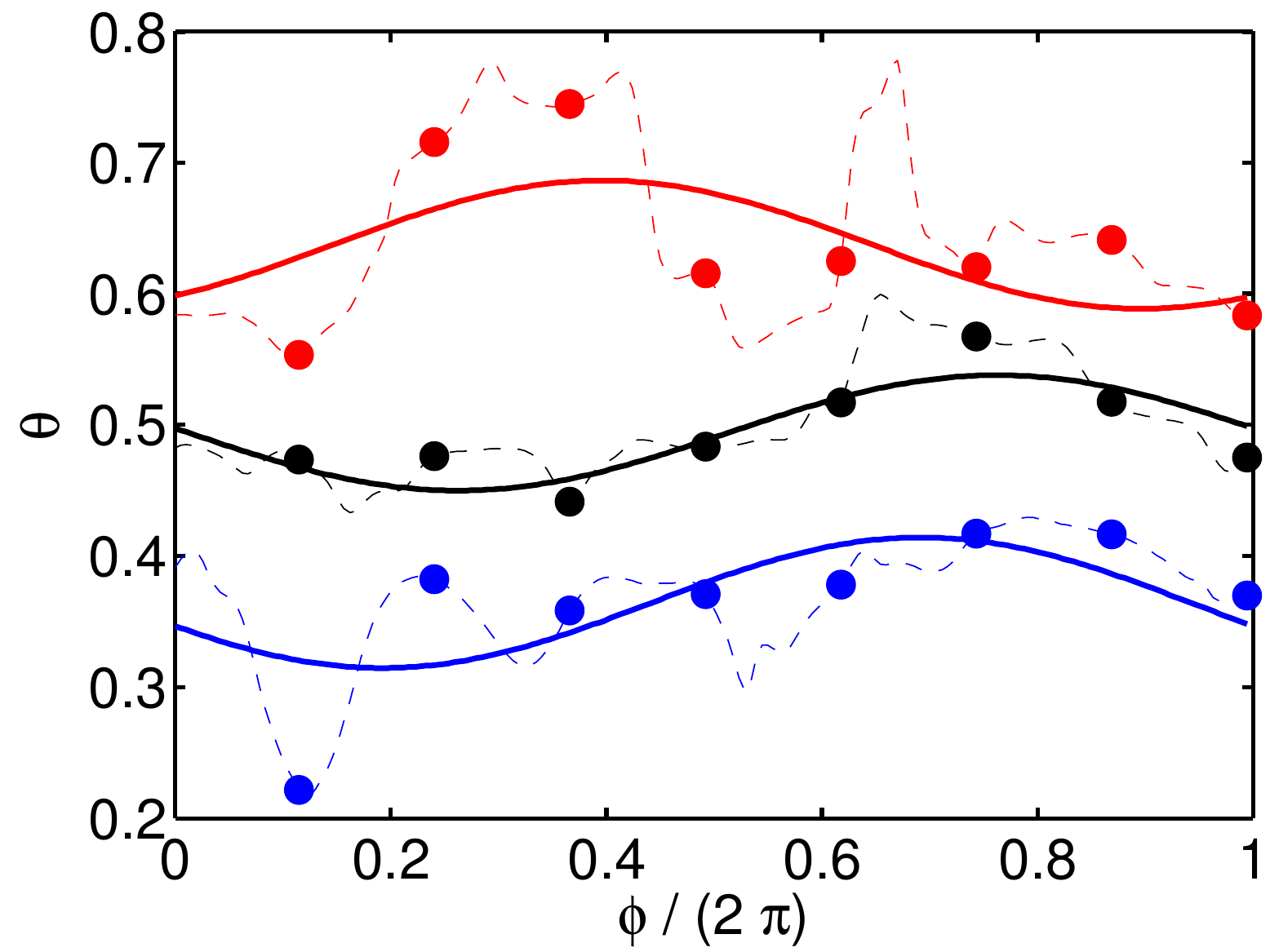}
  \caption{Red (top), black (middle), and blue (bottom) data indicate the azimuthal temperature profile at the heights $0.25L$, $0.50L$, and $0.75L$, respectively, for the flow state indicated in figure \ref{Fig_S10_rollstate2}b. The dots indicate the data of $8$ azimuthally equally spaced probes per horizontal plane. Based on the cosine fits (solid lines) obtained from this probe data this state would be identified as a DRS, while figure \ref{Fig_S10_rollstate2}b shows that it is in fact a SRS with a strong corner flow. The dashed lines show that a higher azimuthal resolution can resolve the double peak structure of the corner flow and the main flow. The temperatures at $0.25L$ ($0.75L$) have been raised (lowered) by $0.1$ for clarity.
  }
  \label{Fig_S09b_Profiles}
\end{figure}

Furthermore, and this is the main issue here, for this relatively low Ra sometimes the data of three rows of $8$ thermistors can give misleading hints for the actual flow state of the system due to the effect of passing plumes and corner rolls on the azimuthal profiles. The importance of these effects is shown in figure \ref{Fig_S09b_Profiles}. This figure shows that based on the data of $8$ azimuthally equally spaced probes at each horizontal level the flow structure shown in figure \ref{Fig_S10_rollstate2}b, which is a SRS with a strong corner flow, would be identified as a DRS. In addition,
 the figure shows that the full azimuthal temperature profiles reveal that a higher azimuthal resolution (e.g., $64$ instead of only $8$ probes) could resolve the double peak structure (originating from  the coexistence of the SRS with some corner flow), which makes it possible to distinguish this state from a real DRS. 

We note that the experiments of Weiss \& Ahlers  \cite{wei10b} show that for $Pr=4.38$ and $Ra \lesssim 3\times 10^9$ the flow state is undefined for about $50\%$ of the time. In the experiments the SRS only establishes itself for higher $Ra$, where the flow state is much better defined, as the 
area of the
corner flow becomes smaller and smaller \cite{sug10}.
It is therefore likely that the examples presented here are primarily important in the low $Ra$ number regime investigated here and are much less common in the higher $Ra$ number regime considered in most experiments. We stress however that one does not know a priori whether one is in a difficult case or in a case where the method works fine. Therefore we feel that it is important to realize that there are some limitations in the methods to determine the flow state from three rows of $8$ equally spaced thermistors. An example of such a difficult case is the breakdown of the LSC when a strong rotation is applied around the cylinder axis \cite{zho09b,ste09}.
 
\section{The relative LSC strength} \label{sec_relativeLSCstrength}
The instantaneous temperature profiles in figure \ref{Fig_S09b_Profiles} show that the LSC is not always clearly present, meaning that the cosine fit becomes poor.
 In order to study this in more detail, we introduce the concept of the relative LSC strength, which we define as the strength of the LSC, i.e. the cosine mode, with respect to the strength of the plumes and turbulent fluctuations, i.e. the fluctuations around the cosine fit.

A convenient way to define a relative LSC strength that is independent of the number of probes is to determine the energy in the different Fourier modes of the azimuthal temperature or the azimuthal vertical velocity profile. In order to obtain a relative LSC strength $\bar{S}_k$ with a number between $0$ and $1$ we normalize the energy in the first Fourier mode, i.e.\  the cosine mode, as follows:
\begin{equation}\label{Eq Relative Strength LSC}
    \bar{S}_k = \mbox{Max}\left( \left( \frac{\sum_{t_b}^{t_e} E_1}{\sum_{t_b}^{t_e} E_{tot}} - \frac{1}{N}\right) / \left(1-\frac{1}{N}\right), 0 \right).
\end{equation}
Here 
$\sum_{t_b}^{t_e} E_1$ indicates the sum of the energy in the first Fourier mode over time, i.e. from the beginning of the simulation $t=t_b$ to the end of the simulation $t=t_e$, $\sum_{t_b}^{t_e} E_{tot}$ the sum of the total energy in all Fourier modes over time, and $N$ the total number of Fourier modes that can be determined. Note that the factor $1/N$ is only used to subtract the part of energy coming from the equi-distribution among all the modes. The number of Fourier modes $N$ that can be determined is $n_p/2$, where $n_p=2^i$ is the number of probes and $i$ is an integer. The subscript $k$ ($=b,m,t$) indicates the height level in the sample at which the relative LSC strength is determined.

Here we will only determine $\bar S_k$  for the middle row of thermistors, thus only $S_m$.
We note that the LSC amplitude $\delta_m$ and the phase $\phi_m$ of the first Fourier mode are identical to the ones determined when 
 using the least square fit in equation (\ref{Eq cosine fit}).
We moreover note that the quantity $\bar S_k$ is {\it not} time resolved and thus does not allow
to study the switching between the SRS and DRS \cite{wei10b}. Results on time-resolved 
relative LSC strengths will be presented in the next section.

The normalization in eq.\ (\ref{Eq Relative Strength LSC}) has been chosen such that
the relative LSC strength $S_m$ always 
has a value between $0$ and $1$: The value  $1$ indicates that the azimuthal profile is a pure 
cosine profile, which is a signature of the LSC according to \cite{bro06}, and the value
$0$ indicates that the magnitude of 
the cosine mode is equal to (or weaker than) the value expected from a random noise 
signal.
$\bar{S}_m \gg 0.5$ indicates that the  cosine fit on average is a
 reasonable approximation of the data, as then 
most energy in the signal is in the first Fourier mode, i.e.\ the cosine mode.  
In contrast, $\bar{S}_m \ll 0.5$  indicates that most energy is in the higher Fourier modes, 
meaning 
 that the application of a cosine fit to the data becomes questionable. 
Somehow arbitrary, we define the SRS as  dominant once  $\bar{S}_m > 0.5$. 

\begin{figure}
  \centering
  \subfigure{\includegraphics[width=3.25in]{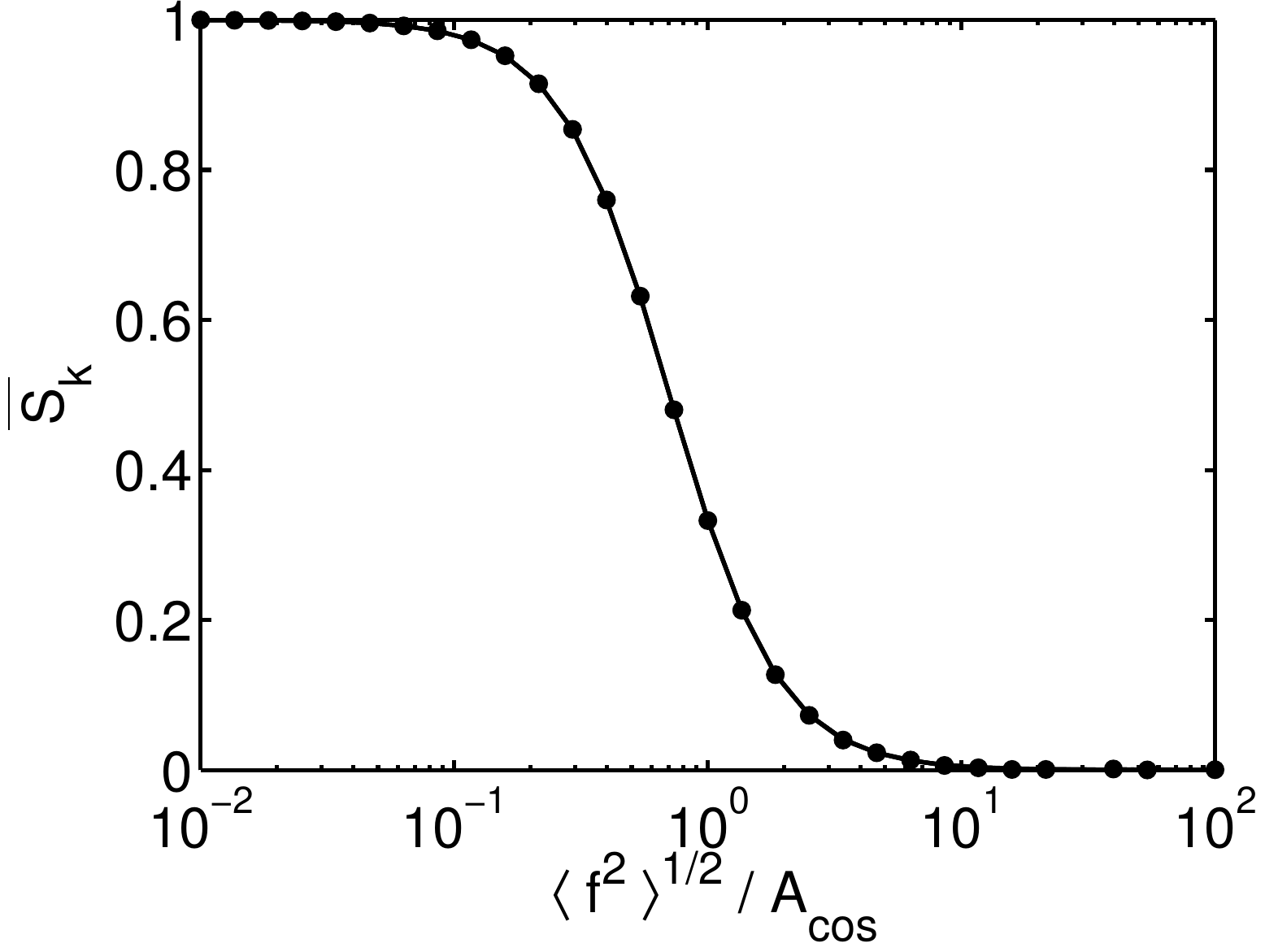}}
	  \caption{The relative strength of the LSC $\bar{S}_k$ versus the relative noise level
$\langle f^2\rangle^{1/2}/A_{cos}$,  calculated from the stochastic model  (equation (\ref{Eq cosine fit2})).
	  For strong  noise $\bar S_k < 0.5$ and the SRS is defined to break down.  
	  }
  \label{Fig_S10_relativestrengthLSC}
\end{figure}

To further demonstrate these properties of $\bar S_k$, we analyze 
the test function
\begin{equation}\label{Eq cosine fit2}
    \theta_i (j)= A_{cos}\cos(\phi_i+j/100)+f(j).
\end{equation}
Here $\phi_i=2 i\pi/n_p$ and $f(j)$ is a set of Gaussian distributed random numbers
 with a certain standard deviation $\langle f^2\rangle^{1/2}$,
the "time" $j$ is varied between $1$ and $125000$, and  
the noise is supposed 
 to model the effect of turbulent fluctuations around the cosine fit. 
The result for the relative strength $\bar S_k$ 
of the LSC as function of the noise level is shown in
 figure \ref{Fig_S10_relativestrengthLSC}. 
Indeed, $1 \ge \bar S_k \ge 0$ and $\bar S_k >0.5$ only if the noise is smaller than the amplitude,
$\left< f^2 \right>^{1/2} /A_{cos} \ll 1$.

We now come back to the numerical data of our simulations.
For each simulation we calculated the relative LSC strength $\bar{S}_m$ based on the instantaneous azimuthal vertical velocity profiles at midheight based on the data of 
either only $8$ or all $64$ probes at the horizontal midplane. We show the results for two aspect ratios, two Rayleigh numbers, and  two Prandtl numbers 
in table \ref{table1}.  For $\Gamma = 1$ the flow is clearly in the SRS, as indicated by $\bar{S}_m \ge 0.65$. For $\Gamma = 1/2$ the SRS is less dominant, in particular
for the large Prandtl number case $Pr=6.4$, where the SRS hardly occurs. This confirms our conclusions of the previous section where we stated that the SRS is not always present in the $\Gamma=1/2$ sample. 

\begin{figure}
  \centering
  \includegraphics[width=3.25in]{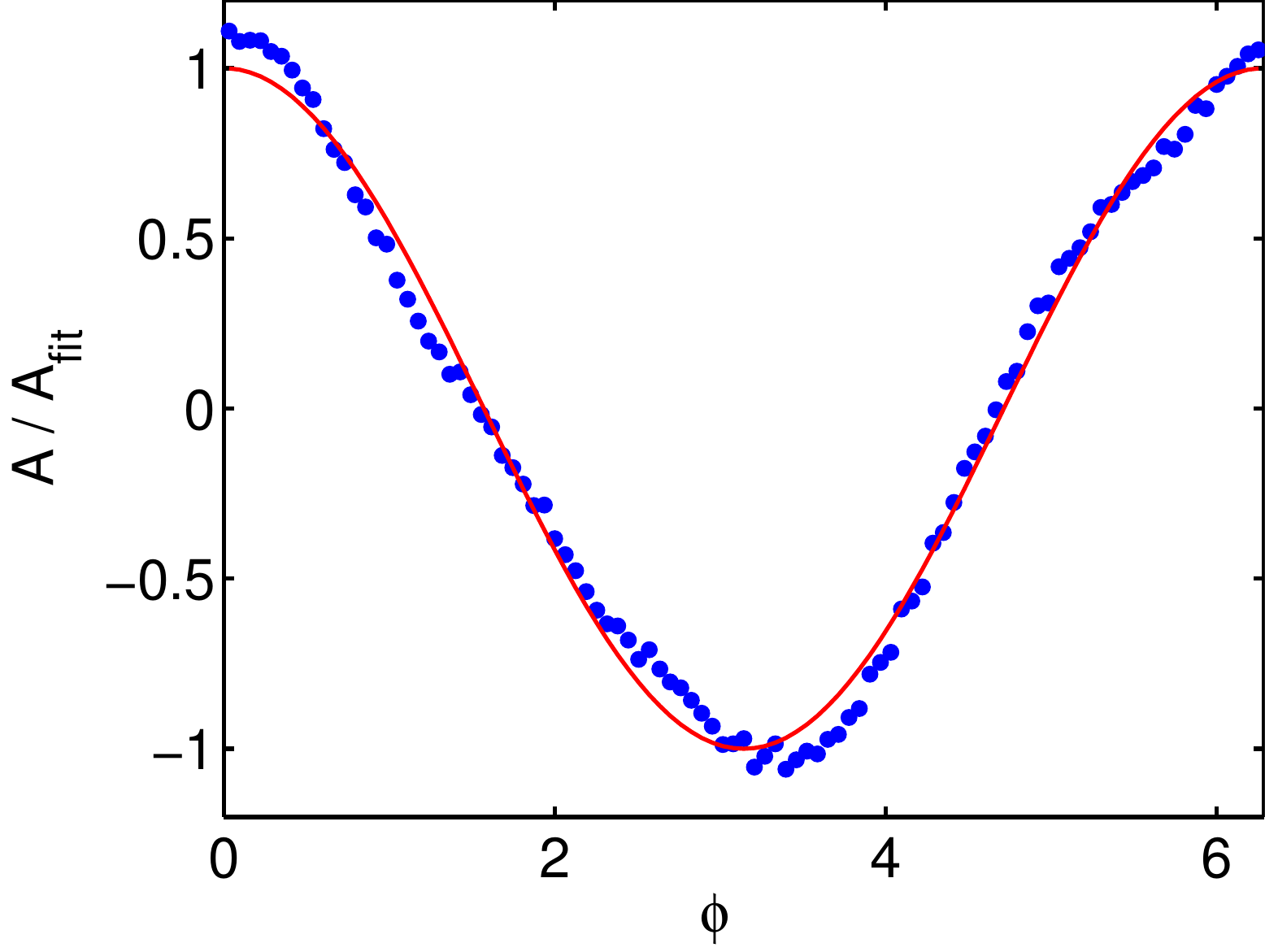}
  \caption{The azimuthal vertical velocity profile after averaging with respect to the orientation of the LSC, which was obtained using a cosine fit, is indicated by the dots. The method has been applied to the data of all $64$ probes in the horizontal midplane obtained in the simulation at $Pr=6.4$, $Ra=1\times 10^8$, and  $\Gamma=1/2$. The solid line indicates a pure cosine profile. }
  \label{Fig_S10_cosineaverage}
\end{figure}

Another method used in the literature \cite{bro06,wei10b} to identify the flow pattern is to average the azimuthal profile with respect to the LSC orientation. This method assumes that the deviations from the cosine fit are due to turbulent fluctuations. It
will work for cases for which  $\bar{S}_m$ is large. 
However, when $\bar{S}_m$ is low, this method gives misleading results as is shown in figure \ref{Fig_S10_cosineaverage} where we apply it  to above discussed data for $Pr=6.4$ in the $\Gamma=1/2$ sample, where the relative LSC strength using the data of all $64$ probes in the horizontal level is  $\bar{S}_m = 0.27$ (see table \ref{table1}): The  average profile obtained with the method of refs.\ \cite{bro06,wei10b} cannot be distinguished from the ones obtained from the cases where $\bar{S}_m$ is larger and therefore the method cannot be used to
claim the dominance of the SRS. 

\begin{figure}
  \centering
  \subfigure{\includegraphics[width=3.25in]{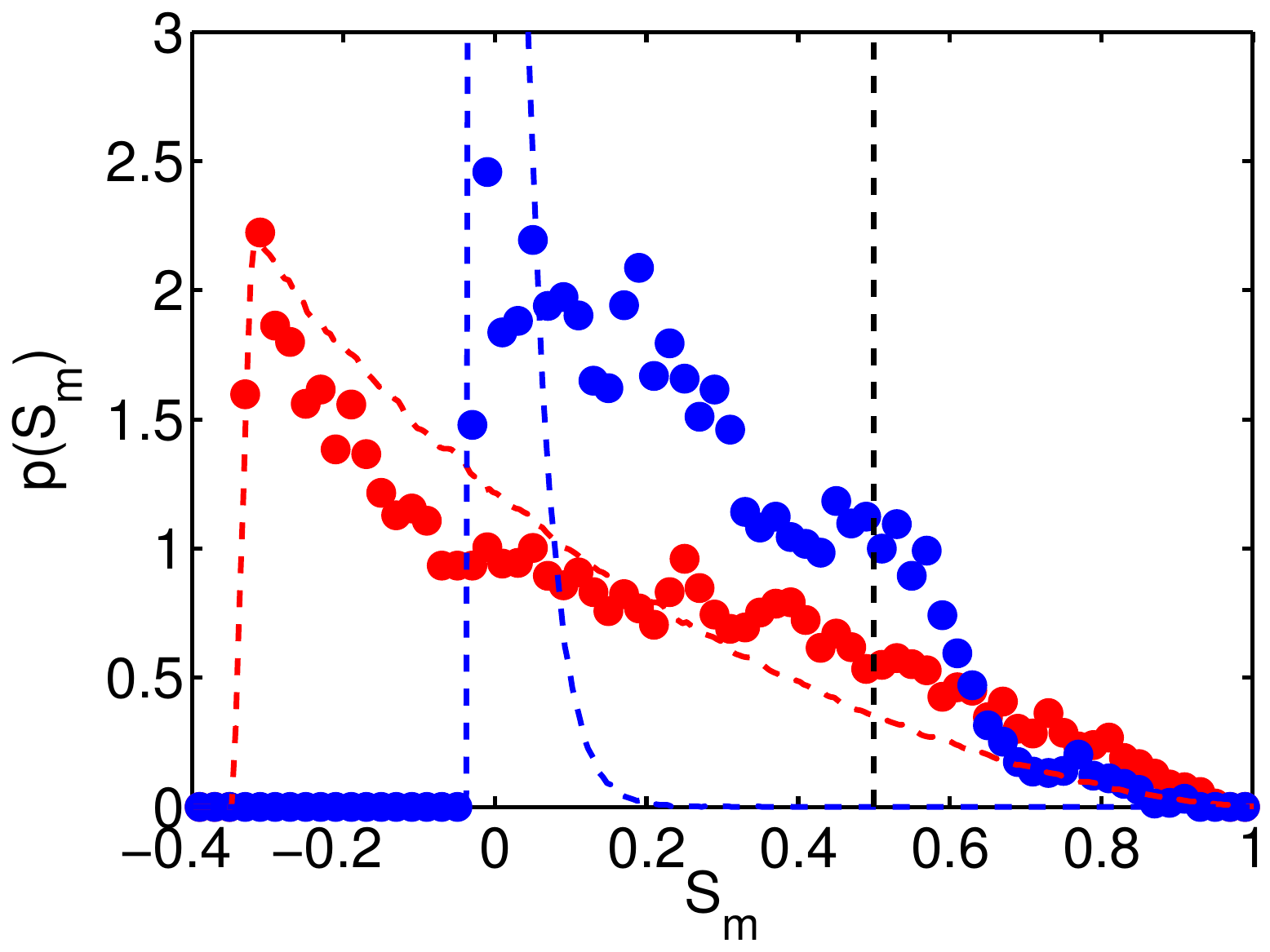}}
  \caption{PDFs of the relative LSC strength at midheight for $Ra=1\times10^8$, $Pr=6.4$, and $\Gamma=0.5$. The red ($-{1\over 3} \le S_m(t) \le 1$) and blue ($-{1\over 31} \le S_m(t) \le 1$) data points indicate the PDF based on the azimuthal vertical velocity profile sampled with only $8$ and all $64$ probes, respectively. The colored dashed lines give the corresponding PDFs based on a purely random signal.
The vertical dashed lines at $S_m=0.5$ indicates the region $0.5 \leq S_m\leq 1.0$ where the first Fourier modes contains at least $50\%$ of the energy of the azimuthal vertical velocity profile.}
  \label{Fig_S10_PDF_S_m}
\end{figure}

\section{Time resolved relative LSC strength} \label{sec_relativeLSCstrength2}
The relative LSC strength can also be calculated on instantaneous azimuthal temperature and vertical velocity profiles. We demonstrate this by constructing the PDF of the relative LSC strength at midheight $S_m(t)$ for instantaneous azimuthal temperature and vertical velocity profiles using the definition 
\begin{equation}\label{Eq Relative Strength LSC2}
   {S}_m (t) =  \left(\frac{E_1(t)}{E_{tot}(t)} - \frac{1}{N}\right) / \left(1-\frac{1}{N}\right).
\end{equation}
The only difference with respect to equation (\ref{Eq Relative Strength LSC}) 
is that $E_1(t)$ and $E_{tot} (t) $ and thus $S_m(t)$ are time dependent and 
that 
we have dropped the criterion that the minimal value should be $0$. 
The reason for the latter
 is that we want to prevent strange jumps in the PDFs of $S_m$ at 0. 
For $8$ probes we have $-{1\over 3} \le S_m(t) \le 1$ and 
for $64$ probes we have $-{1\over 31} \le S_m(t) \le 1$. Note that these intervals follow directly from equation \ref{Eq Relative Strength LSC2} by filling in the limiting cases, i.e.\ ${E_1(t)} / E_{tot}(t)=0$ and $E_1(t) / E_{tot}(t)=1$, and by using the relation $N=n_p/2$, where $n_p$ is the number of probes, see previous section.

Figure \ref{Fig_S10_PDF_S_m} shows the PDFs for $S_m$ based on the measurement of the vertical velocity by only $8$ or all $64$ equally spaced probes at midheight in the simulation with 
 $Ra=1\times10^8$, $Pr=6.4$, and  in the  $\Gamma=0.5$ sample. In addition, for both cases 
 we determined the distribution of $S_m(t)$ for a  random signal. 
When we compare the PDFs based on the simulation data with the distributions obtained for random signals,
 we see that the difference is relatively small when the data of only $8$ probes is considered, 
whereas a much larger difference is obtained
 when the data of all $64$ probes is taken into account. 
Figure \ref{Fig_S10_PDF_S_m} thus clearly shows that it can be beneficial to use
 more than $8$ azimuthally equally spaced probes, 
since it becomes much easier to show that certain events are statistically relevant.

\section{Conclusions} \label{sec_conclusions}
We studied the LSC dynamics in DNS simulations by investigating the azimuthal temperature and vertical velocity profiles obtained from $64$ equally spaced numerical probes at three different heights. For $Pr=6.4$ in a $\Gamma=1$ sample we find that the azimuthal profile is well presented with the data of $8$ numerical probes, a number normally used in experiments, as the LSC orientation obtained by a cosine fit is the same when the data of $8$ and $64$ probes is considered. We find that the improved
 azimuthal resolution (64 instead of 8 probes in the sidewall at one height)  can reveal
 the effect
 of the plumes. 
 
In agreement with the findings of Xi and Xia \cite{xi08} and Weiss and Ahlers  \cite{wei10b} 
we find that there is more disorder present in the
  $\Gamma=1/2$ sample than in the 
 $\Gamma=1$ sample.
For the $\Gamma = 1/2$ case we also show that when the azimuthal temperature profile is only determined from $8$ probes per horizontal level a SRS can erroneously be identified as a DRS, because the azimuthal resolution is too small to distinguish between 
the structure of the corner flow and the main roll. Here we again stress that this result is obtained in the relatively low $Ra$ number regime. The experimental results of 
Weiss \& Ahlers  \cite{wei10b} show that for the low $Ra$ number regime the flow state is undefined for about $50\%$ of the time.
 For $\Gamma = 1/2$ the SRS only establishes itself for higher $Ra$, where the flow state is much better defined \cite{wei10b}. 
It is therefore likely that the examples presented here are primarily important in this low $Ra$ number regime investigated here and are much less common in the higher $Ra$ number regime.

We quantified the LSC strength  relative  to the turbulent fluctuations by determining the ratio between the energy in the first Fourier mode and the energy in all Fourier modes of the azimuthal temperature and azimuthal vertical velocity profiles. We find that the relative LSC strength at $Ra=1\times10^8$ is considerably lower in the  $\Gamma=1/2$ sample than in the $\Gamma=1$ sample, 
i.e.\ that the SRS is much less pronounced in the  $\Gamma=1/2$ sample than in the $\Gamma=1$ sample. This determination of the relative LSC strength can be applied directly to available experimental data to determine whether the SRS is present in high Ra number thermal convection and in rotating RB convection. 

{\it Acknowledgements:} We gratefully acknowledge various discussions with Guenter Ahlers over this line of research and his helpful comments on our manuscript. We also acknowledge discussions with Stefan Weiss and Eric Brown. We thank R.\ Verzicco for providing us with the numerical code. The work is supported by the Foundation for Fundamental Research on Matter (FOM) and the National Computing Facilities (NCF), both sponsored by NWO. The computations have been performed on the Huygens supercomputer of SARA in Amsterdam.

\end{document}